\documentclass[a4paper, 11pt]{article}

\usepackage{jcappub}
\usepackage[top=25truemm, bottom=30truemm, left=20truemm, right=20truemm]{geometry}

\usepackage{graphicx}
\usepackage{bm,latexsym,amsmath,amssymb,amsfonts,mathrsfs}
\usepackage{float}
\usepackage{xcolor}
\usepackage{physics}
\usepackage{color}
\usepackage{comment}
\usepackage{here}
\usepackage[normalem]{ulem}
\usepackage{empheq}

\allowdisplaybreaks[4]

\title{%
    Primordial black holes in excursion set theory:\\%
    Formation probabilities, mass functions,\\%
    and window functions%
}%
\author[a]{Hayami Iizuka,}
\author[b]{Daiki Saito,}
\author[c]{and Koki Tokeshi}

\emailAdd{h.iizuka@rikkyo.ac.jp}
\emailAdd{dsaito@ewha.ac.kr}
\emailAdd{koki.tokeshi@phys.ens.fr}

\affiliation[a]{%
    \textit{Department of Physics, Rikkyo University, Toshima, Tokyo 171-8501, Japan}
}%
\affiliation[b]{%
    \textit{Department of Science Education, Ewha Womans University, Seoul 03760, Korea}
}%
\affiliation[c]{%
    \textit{Laboratoire de Physique de l'\'{E}cole Normale Sup\'{e}rieure, ENS, CNRS,\\ 
Universit\'{e} PSL, Sorbonne Universit\'{e}, Universit\'{e} Paris Cit\'{e}, 75005 Paris, France}
}%

\date{\today}
\abstract{
We study the mass function of primordial black holes (PBHs) within the excursion-set theory, in which the response of the stochastic density contrast to the variation of the coarse-graining scale is described by colored noises. 
For several window functions often used in the literature, we investigate how this choice affects the formation probability as well as the resultant mass function of PBHs. 
It is found that the low-mass tail of the mass function differs from the one predicted from Carr's formula. 
The difference comes from the prevalence of correlated noises, by which degeneracy of the formation probabilities ceases to exist. 
Nevertheless, Carr's formula still provides a practical estimation in the vicinity of the characteristic mass scale, as long as a smooth window function in Fourier-space is used. 
}

\begin{document}
\flushbottom
\maketitle

\section{Introduction}
Primordial black holes~(PBHs) are hypothetical black holes that may have formed in the early Universe~\cite{Zeldovich:1967lct, Hawking:1971ei, Carr:1974nx}. 
Since their original proposal more than five decades ago, PBHs have continued to attract attention as probes of the early Universe, in connection with dark matter~\cite{Chapline:1975ojl, Carr:2016drx, Carr:2021bzv}, gravitational-wave observations~\cite{Bird:2016dcv, Clesse:2016vqa, Sasaki:2016jop, Sasaki:2018dmp}, and the origin of supermassive black holes~\cite{Carr:2018rid}. 
Though there is no direct evidence for the existence of PBHs, their abundance has been studied through various formulations and/or methods, ranging from Carr's pioneering cumulative formula~\cite{Carr:1975qj} to those based on numerical simulations of gravitational collapse~\cite{Niemeyer:1999ak,Shibata:1999yda}.
Meanwhile, observational constraints over a broad range of masses already put bounds on the viable PBH abundance from above~\cite{Carr:2009jm, Carr:2020gox}, due to which reliable theoretical predictions of the PBH abundance and mass function become increasingly important; see, for instance, Refs.~\cite{Yoo:2022mzl, Escriva:2022duf, Carr:2026hot} for recent reviews. 

The most widely studied formation mechanism of PBHs is based on an enhancement of primordial perturbations by an underlying inflationary model, which could lead to gravitational collapse of the associated overdense regions during the radiation-dominated era. 
Once the power spectrum of the curvature perturbation is provided or assumed, a standard estimate of the abundance of PBHs found in the literature follows from the probability that the coarse-grained density contrast exceeds a certain threshold at a given coarse-graining scale, which is related to the mass of the resultant PBH. 
This method is easy to handle and may sometimes be reliable, as we will see, however for instance the formation process in which a smaller PBH is absorbed by a larger one is not taken into account, leading to an underestimation of the abundance. 
Among other formulas such as the statistics of peaks~\cite{Bardeen:1985tr}, the excursion-set method~\cite{Bond:1990iw} resolves this issue, on which the present article focuses. 

The excursion-set theory was originally developed to describe structure formation in terms of the stochastic process~\cite{Chandrasekhar:1943ws} of a coarse-grained density contrast as the coarse-graining scale is varied~\cite{Bond:1990iw, Nikakhtar:2018qqg} (see Ref.~\cite{Zentner:2006vw} for a review) and there are several works~\cite{Green:1999xm, MoradinezhadDizgah:2019wjf, DeLuca:2020ioi, Auclair:2020csm, Erfani:2021rmw, Auclair:2024jwj, Kushwaha:2025zpz, Saito:2025sny, Auclair:2026tfy} in which it was applied to the context of PBHs. 
In this framework, formation of PBHs is described by the first crossing of a threshold of the coarse-grained density contrast, rather than by the probability that it exceeds the threshold at a fixed scale. 
Therefore, the abundance, as well as the formation probability itself, estimated within the excursion-set method may not coincide with the estimation based on Carr's pioneering formula~\cite{Carr:1975qj}.
Since the stochastic process that determines the mass function of PBHs is controlled by the statistical profile of the primordial curvature perturbation and the choice of the window function, which coarse-grains the density contrast, it is necessary to examine the effect of those uncertainties, \textit{i.e.}, in what ways and how much 
the resultant mass function of PBHs depends on those functional forms. 

Our particular interest within the excursion-set method is twofold.
One is related to the $\mathcal{O} (1)$-factor discrepancy from Carr's cumulative formula, in connection with the so-called cloud-in-cloud issue. 
In the original (\textit{i.e.}~non-PBH) context and in the highly idealized situation, the excursion-set method resolves the issue, giving the ``factor two'' due to the absence of the correlation across scales. 
This no longer holds once the method is extended for estimating the formation probabilities of PBHs, because even in the idealized case, correlation across scales arises, as recently pointed out and demonstrated for the discontinuous window function in Fourier space in Ref.~\cite{Kushwaha:2025zpz} (see Ref.~\cite{Auclair:2026tfy} for a different treatment in which the noise is taken as uncorrelated). 
In such cases, the $\mathcal{O} (1)$-factor deviates from two. 
In addition, the situation becomes more involved once a smooth window function is implemented, which is the other interest of the present article. 
A demonstration using the Gaussian window function is presented in Ref.~\cite{Saito:2025sny}, to find that the degeneracy of the two kinds of formation probabilities that we introduce in Section~\ref{subsec:fprob} vanishes. 
This outcome is expected to affect the resultant abundance of PBHs, and thus, we are motivated to clarify how the choice of the window functions influences the mass function, especially around the characteristic scale at which it is maximized. 

In this article, we study the formation of PBHs within the excursion-set method by numerically generating stochastic trajectories of the coarse-grained density contrast, whose response to the variation of the coarse-graining scale is driven by the correlated, or colored, noises. 
The formation probabilities are extracted from those trajectories, which are then converted into the mass function by extending the analysis of Ref.~\cite{Saito:2025sny} that restricted itself to the probabilities and to the Fourier-space top-hat and the Gaussian window functions. 
We also clarify the dependence of the numerically reconstructed mass function of PBHs on the choice of the window function. 
On this point, particular attention is paid to the real-space top-hat window function, which has not been studied in the previous works~\cite{Kushwaha:2025zpz, Saito:2025sny} for simplicity, or due to the technical difficulty that we will discuss. 

The rest of this article is organized as follows. 
In Section~\ref{formulation}, we introduce the coarse-grained density contrast, fix the localized Gaussian power spectrum to be used throughout the article, and review the window functions that are often used in the literature. 
The covariance matrix and the variance of the stochastic noise that drives the excursion-set trajectory are derived analytically, though for the real-space top-hat window function, we retain the integral expression. 
Based on these preparatory materials, in Section~\ref{stochastic}, we numerically solve the stochastic process exhibited by the coarse-grained density contrast, extracting the formation probabilities. 
We then reconstruct the mass function in Section~\ref{massfunction}, comparing it with the estimate based on Carr's formula. 
We also investigate the dependence of the mass function on the choice of the window function. 
Finally, Section~\ref{conclusion} is devoted to summarizing our results. 

Throughout this article, we use units in which both the speed of light and Newton's gravitational constant are set to unity, $c=1~\text{and}~G=1$.

\section{Formulation}
\label{formulation}

This section provides the basic equations, which serve as a foundation for numerical simulations of the stochastic process in Section~\ref{stochastic}. 
Section~\ref{sec:density} introduces the coarse-grained density contrast and the stochastic process that it obeys. 
The statistical properties of the correlated noises are also reviewed, giving the general formula in which concrete forms of the power spectrum and window function are kept unspecified. 
Given that the choice of the window function does matter and is thus one of our interest, its feature is discussed in Section~\ref{sec:pw}, where the localized Gaussian power spectrum is fixed. 
The concrete expressions for the covariance and the variance of the noise are given in Section~\ref{subsec:ncov}. 

\subsection{Coarse-grained density contrast in excursion-set method}
\label{sec:density}

At leading order of the gradient expansion, and in the radiation-dominated era, the density contrast and the conserved curvature perturbation are related as~\cite{Harada:2015yda}, 
\begin{equation}
    \delta( t, \, \mathbf{x}) 
    \simeq 
    - \frac{8}{9} 
    \exp \qty( 
        -\frac{5}{2} \zeta
    ) 
    \qty(\frac{1}{aH})^{2} 
    \Delta
    \exp \qty( \frac{\zeta}{2} ) 
    \simeq 
    - \frac{4}{9} 
    \qty( \frac{1}{aH} )^{2} 
    \Delta \zeta 
    \,\, ,
\end{equation}
on a comoving slice with $aH$ evaluated at the time of interest. 
In the second equality, we have performed a linear approximation, and this linearity will be assumed throughout. 
We also assume the Gaussianity of the curvature perturbation $\widetilde{\zeta} (\mathbf{k})$, together with its vanishing mean, $\expval*{ \widetilde{\zeta} (\mathbf{k}) } = 0$.
Then, the density contrast is in turn a Gaussian variable.
In Fourier space, the linearized relation reads 
\begin{equation}
    \widetilde{\delta} (\mathbf{k}, \, t) 
    = 
    \frac{4}{9} \qty(
        \frac{k}{aH} 
    )^{2} 
    \widetilde{\zeta} (\mathbf{k}) 
    \,\, .
\end{equation}

Since in most cases a PBH may be formed when the relevant mode of the super-horizon perturbation reenters the horizon, the coarse-graining scale is identified through $\tau \equiv aH$. 
With this notation, we define the coarse-grained density contrast at scale $\tau$ as 
\begin{equation}
    \delta(\tau, \, \mathbf{x})
    \equiv 
    \frac{4}{9}
    \int \frac{d^{3} k}{(2 \pi)^{3}} 
    \, 
    e^{i \mathbf{k} \cdot \mathbf{x}} 
    \, 
    \widetilde{\zeta} (\mathbf{k})
    \eval{ 
        [ z^{2} \widetilde{W} (z) ] 
    }_{z = k / \tau} 
    \,\, , 
    \label{eq:cgdel}
\end{equation}
where $\widetilde{W} (z)$ is a given window function in Fourier space. 
Given that the excursion-set method monitors the variation of $\delta(\tau, \, \mathbf{x})$, with respect to the variation of the coarse-graining scale $1 / \tau$ at a fixed spatial location, from this point onward, we suppress the explicit spatial dependence and focus on a fixed reference position. 
We write $\delta (\tau, \, \mathbf{x}$) simply as $\delta(\tau)$ accordingly. 
The derivative of Eq.~(\ref{eq:cgdel}) with respect to $\tau$ gives 
\begin{equation}
    \frac{\partial \delta (\tau)}{\partial \tau}
    = 
    - \frac{4}{9}
    \frac{1}{\tau^{2}}
    \int \frac{\dd^{3} k}{(2 \pi)^{3}} 
    \, 
    e^{i \mathbf{k} \cdot \mathbf{x}}
    \, k \, 
    \widetilde \zeta(\mathbf{k})
    \frac{\dd}{\dd z}
    \eval{ 
        [ z^{2} \widetilde{W} (z) ]
    }_{z = k / \tau} 
    \,\, .
\label{eq:deriv_delta}
\end{equation}
Since the curvature perturbation is a stochastic quantity, the right-hand side of Eq.~(\ref{eq:deriv_delta}) may be regarded as a stochastic noise that drives the stochastic process of $\delta (\tau)$ with respect to the variation of the coarse-graining scale $1 / \tau$, governed by Eq.~(\ref{eq:deriv_delta}). 
This motivates us to define 
the stochastic noise $\xi (\tau)$ by the right-hand side of Eq.~(\ref{eq:deriv_delta}), so that the stochastic process is defined by 
\begin{equation}
    \pdv{\delta(\tau)}{\tau} 
    = 
    \xi(\tau) 
    \,\, . 
    \label{eq:stochastic}
\end{equation}
This stochasticity should not be interpreted as dynamical time evolution, but rather as the random scale dependence of the coarse-grained density contrast at a fixed point in space. 
The statistical properties of the noise $\xi (\tau)$ are then determined by the covariance, which is given by 
\begin{align}
    \langle 
        \xi(\tau_1) \xi(\tau_2) 
    \rangle
    &=
    \frac{16}{81}
    \frac{1}{\tau^{2}_{1} \tau^{2}_{2}}
    \int \frac{\dd^{3} k_{1}}{(2 \pi)^{3}}
    \int \frac{\dd^{3} k_{2}}{(2 \pi)^{3}} 
    \, 
    e^{i (\mathbf{k}_{1} + \mathbf{k}_{2}) \cdot \mathbf{x}} 
    \, k_{1} k_{2} \, 
    \langle 
        \widetilde{\zeta} (\mathbf{k}_{1}) 
        \widetilde{\zeta} (\mathbf{k}_{2}) 
    \rangle 
    \notag \\ 
    &\quad \times 
    \eval{ 
        \frac{\dd}{\dd z_{1}}
        [
            z_{1}^{2} \widetilde{W} (z_{1}) 
        ] 
    }_{z_{1} = k / \tau_{1}} 
    \eval{ 
        \frac{\dd}{\dd z_{2}} 
        [
            z_{2}^{2} \widetilde{W} (z_{2}) 
        ] 
    }_{z_{2} = k / \tau_{2}} 
    \notag \\ 
    &= 
    \frac{16}{81} 
        \frac{1}{\tau_{1}^{2} \tau_{2}^{2}} 
        \int_{0}^{\infty} \frac{\dd k}{k} \, 
        \mathcal{P}_{\zeta} (k) 
        k^{2} 
        \qty{ 
            \dv{z_{1}} 
            \qty[ 
                z_{1}^{2} \widetilde{W} (z_{1}) 
            ]
        }_{z_{1} = k / \tau_{1}} 
        \qty{ 
            \dv{z_{2}} 
            \qty[ 
                z_{2}^{2} \widetilde{W} (z_{2}) 
            ]
        }_{z_{2} = k / \tau_{2}} 
        \,\, , 
    \label{eq:covariancexi}
\end{align}
where the non-dimensionalized power spectrum of the curvature perturbation $\mathcal P_{\zeta}(k)$ is defined through 
\begin{equation*}
    \expval*{ 
        \widetilde{\zeta} (\mathbf{k}_1) 
        \widetilde{\zeta} (\mathbf{k}_2) 
    } 
    = (2 \pi)^{3}
    \delta_{\rm D} (\mathbf{k}_{1} + \mathbf{k}_{2}) 
    \frac{2 \pi^{2}}{k_{1}^{3}}
    \mathcal{P}_{\zeta} (k_{1}) 
    \,\, .
\end{equation*}
From Eq.~(\ref{eq:covariancexi}), it can be seen that there are, in general, correlations among noises across different scales, that is, the noises are colored, even for the highly idealized situation with the Fourier-space top-hat window function. 
This colored nature of the noises originates from the fact that Eq.~(\ref{eq:covariancexi}) contains the window function without the derivative, as opposed to the excursion-set method for halo formation~\cite{Bond:1990iw}.
We note that, under another choice of the time variable, the excursion-set method in the context of the formation of PBHs can also be reduced to solving the stochastic dynamics with noises being uncorrelated, while the threshold becomes scale-dependent~\cite{Auclair:2026tfy}. 
Nevertheless, in this article, we employ the formalism that retains the correlated nature of the noise, following Refs.~\cite{Kushwaha:2025zpz, Saito:2025sny}. 

Taking an equal-time limit $\tau_{1} \to \tau_{2} = \tau$ in Eq.~(\ref{eq:covariancexi}), one obtains the variance of the noise 
\begin{equation}
    \expval*{ 
        [ \xi(\tau) ]^2 
    } 
    = 
    \frac{16}{81} 
    \int\frac{\dd k}{k} 
    \, 
    \mathcal{P}_\zeta (k) 
    \frac{k^{2}}{\tau^{4}} 
    \left\{
        \frac{\dd}{\dd z}
        \eval{ 
            [
                z^{2} \widetilde{W} (z)
            ] 
        }_{z = k / \tau}
    \right\}^{2} 
    =
    \frac{16}{81} 
    \int_{0}^{\infty} \frac{\dd z}{z} 
    \, 
    \frac{z^{2}}{\tau^{2}} 
    \mathcal{P}_{\zeta} (k = \tau z) 
    \qty{ 
        \dv{z} 
        [ 
            z^{2} \widetilde{W} (z) 
        ] 
    }^{2} 
    \,\, .
    \label{eq:variancexi}
\end{equation}
From Eqs.~(\ref{eq:stochastic}),~(\ref{eq:covariancexi}) and (\ref{eq:variancexi}), we obtain the covariance and variance of $\delta(\tau)$ as follows
\begin{subequations}
    \begin{align}
        \langle 
            \delta (\tau_{1}) \delta(\tau_{2}) 
        \rangle 
        &= 
        \frac{16}{81} 
        \int_{0}^{\infty} \frac{\dd k}{k} 
        \, 
        \mathcal{P}_{\zeta} (k) 
        \qty[ 
            \eval{ 
                z_{1}^{2} 
                \widetilde{W} (z_{1}) 
            }_{z_{1} = k / \tau_{1}} 
        ] 
        \qty[ 
            \eval{ 
                z_{2}^{2} 
                \widetilde{W} (z_{2}) 
            }_{z_{2} = k / \tau_{2}} 
        ] 
    \,\, , 
    \label{eq:covd}
    \\[2.0ex] 
    \expval*{ 
        \qty[\delta(\tau)]^{2} 
    } 
    &= 
    \frac{16}{81} 
    \int_{0}^{\infty} \frac{\dd z}{z} 
    \, 
    \mathcal{P}_{\zeta} (\tau z) 
    \qty[ 
        \eval{ 
            z^{2} \widetilde{W} (z) 
        }_{z=k/\tau} 
    ]^{2} 
    \,\, . 
    \label{eq:vard}
    \end{align}
\end{subequations}
The variances of the noise, Eq.~(\ref{eq:variancexi}), and that of the coarse-grained density contrast, Eq.~(\ref{eq:vard}), are used to confirm the consistency of our numerical simulations that performed in Section~\ref{stochastic}.
The latter has also been used in the literature to estimate the abundance of PBHs, within Carr's seminal formula~\cite{Carr:1975qj}, which will be contrasted in Section~\ref{subsec:fprob} with that obtained from the excursion-set method. 

\subsection{Localized power spectrum and window functions}
\label{sec:pw}

To proceed further, functional forms of both window function $\widetilde{W} (z)$ and power spectrum $\mathcal{P}_{\zeta} (k)$ need to be specified. 
For a scenario that realizes PBHs at small scales and, at the same time, is consistent with large-scale observations, a power spectrum with a localized peak is often assumed. 
Following Ref.~\cite{Saito:2025sny}, we assume the (tilted) Gaussian power spectrum given by 
\begin{equation}
    \mathcal{P}_\zeta(k)
    =
    \frac{\mathcal{P}_0k_\star}{\sqrt{2\pi\Delta^2}}
    \left(\frac{k}{k_\star}\right)^\alpha
    \exp \left[
    -\frac{(k-k_\star)^2}{2\Delta^2}
    \right] 
    \,\, , 
    \label{eq:ps_alpha}
\end{equation}
where $k_\star$ denotes the peak scale, $\Delta$ characterizes the width of the localized enhancement, and $\alpha$ controls the tilt of the spectrum around the peak as well as its infrared behavior.
For our numerical convenience later, we introduce the following non-dimensionalized variables 
\begin{equation}
    \widetilde{\tau}\equiv \frac{\tau}{\Delta},
    \qquad
    \widetilde{k}_*\equiv \frac{k_\star}{\Delta}.
\end{equation}

The variance and covariance of the noise are then given, respectively, by 
\begin{subequations}
\begin{align} 
    \expval*{ 
        [ \Delta \xi (\tau) ]^{2} 
    } 
    &= 
    \frac{16}{81}
    \frac{\mathcal{P}_0}{\sqrt{2\pi}}\,
    \widetilde{k}_*^{\,1-\alpha}\,
    \widetilde\tau^{\,\alpha-2}
    \int_0^\infty \dd z 
    \,
    z^{\alpha+3}
    \exp\left[
        -\frac{(\widetilde{\tau} z-\widetilde{k}_*)^2}{2}
    \right]
    \left[
        2\widetilde W(z)
        +
        z\,\widetilde W'(z)
    \right]^2 
    \,\, , 
    \label{eq:xi_variance_z}
    \\ 
    \Delta^{2} 
    \langle
        \xi(\tau_{1}) \xi(\tau_{2})
    \rangle
    &=
    \frac{16}{81}
    \frac{\mathcal{P}_{0}}{\sqrt{2 \pi}}  
    \frac{\widetilde{k}_{*}^{1 - \alpha}}{\tau_{1}^{2} \tau_{2}^{2}} 
    \int_{0}^{\infty} \frac{\dd x}{x} \,
    x^{\alpha + 2}
    \exp \bigg[ 
        -\frac{ ( x - \widetilde{k}_{*} )^{2} }{2} 
    \bigg] 
    \notag \\ 
    &\quad \qquad\qquad\qquad\qquad \times 
    \left\{
        \frac{\dd}{\dd z_1}
        \left[z_1^2\widetilde W(z_1)\right]
    \right\}_{z_1=x/\widetilde\tau_1}
    \left\{
        \frac{\dd}{\dd z_2}
        \left[z_2^2\widetilde W(z_2)\right]
    \right\}_{z_2=x/\widetilde\tau_2} 
    \,\, , 
    \label{eq:app_xi_cov_general}
\end{align}
\end{subequations}
and the covariance of the coarse-grained density contrast is given by
\begin{equation}
    \langle
        \delta (\tau_{1}) \delta(\tau_{2}) 
    \rangle
    = 
    \frac{16}{81} 
    \frac{\mathcal{P}_{0}}{\sqrt{2 \pi}} 
    \widetilde {k}_{*}^{1 - \alpha} 
    \int_{0}^{\infty} \frac{\dd x}{x} 
    \,
    \exp \bigg[
        - \frac{ (x - \widetilde{k}_{*})^{2} }{2}
    \bigg] 
    \eval{ 
        [ z_{1}^{2} \widetilde{W} (z_{1}) ]
    }_{z_{1} = x / \widetilde\tau_{1}} 
    \eval{ 
        [ z_{2}^{2} \widetilde{W} (z_{2}) ] 
    }_{z_{2} = x / \widetilde\tau_{2}} 
    \,\, . 
    \label{eq:app_delta_cov_general}
\end{equation}
For smooth window functions, the derivatives in Eq.~(\ref{eq:xi_variance_z}) and Eq.~\eqref{eq:app_xi_cov_general} do not yield the Dirac-$\delta$ function, whereas for the Fourier-space top-hat window function, the derivative is distributional that gives rise to the uncorrelated component of the noise. 

\vspace{1.0\baselineskip}

\begin{figure}
    \centering
    \includegraphics[width=0.65\linewidth]{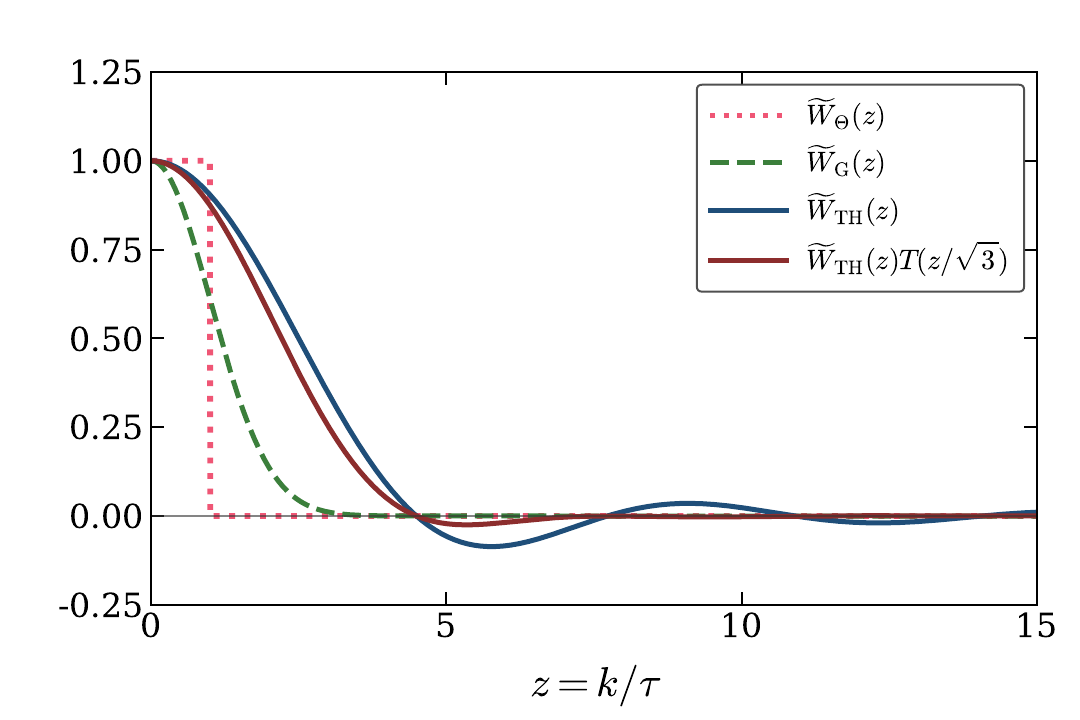}
    \caption{
        The window functions often considered in the literature: Fourier-space top-hat window function (orange dotted), Gaussian window function (green dashed), and real-space top-hat window function (blue solid and wine-red solid).
    }
    \label{fig:wf}
\end{figure}

Having fixed the power spectrum (\ref{eq:ps_alpha}), the remaining thing is to specify a concrete form of the window function to be fully ready for numerical simulations. 
We consider three representative choices as shown in Figure~\ref{fig:wf}. 
\begin{itemize}
    \item 
        Fourier-space top-hat window function: 
        \begin{equation}
            \widetilde W_{\Theta}(z)=\Theta(1-z) 
            \,\, , 
            \label{eq:WF1}
        \end{equation}
    \item 
        Gaussian window function: 
        \begin{equation}
            \widetilde W_{\rm G}(z)=\exp\left(-\frac{z^2}{2}\right)
            \label{eq:WF2}
            \,\, , 
        \end{equation}
    \item 
        Real-space top-hat window function: 
        \begin{equation}
            \widetilde W_{\rm TH}(z)=3\frac{\sin z-z\cos z}{z^3}
                = \frac{3 j_{1} (z)}{z} 
            \label{eq:WF3}
            \,\, . 
        \end{equation}
\end{itemize}
where $j_{1} (z) = (\sin z - z \cos z) / z^{2}$ is the spherical Bessel function of the first kind of order one. 

We here make a few comments on why the three kinds of window functions are considered in what follows. 
In the context of the excursion-set method and PBHs, where the noises are correlated, a demonstration was recently presented in Ref.~\cite{Kushwaha:2025zpz}, focusing on the discontinuous window function~(\ref{eq:WF1}). 
There, though both uncolored and colored noises are present, the driving source for the coarse-grained density contrast to pierce the threshold is restricted to the uncolored noise, as was discussed in Ref.~\cite{Saito:2025sny}. 
The resultant probabilities are thus degenerated near the characteristic scale, contributing to the mass distribution nearly comparably. 
Reference~\cite{Saito:2025sny} studied the formation probabilities of PBHs, based on the formulation in the former reference but with the Gaussian window function, Eq.~(\ref{eq:WF2}), as an example of a smooth coarse-graining. 
There, it was found that the continuity of the window function renders the degeneracy of the formation probabilities to vanish. 
Together with the long-standing uncertainty that the choice of the window function alters the resultant abundance of PBHs~\cite{Ando:2018qdb, Tokeshi:2020tjq, Yoo:2020dkz} in the standard (\textit{i.e.}~non-excursion-set) context, we are therefore interested in the influence of the choice of the window function on the resultant abundance of PBHs, in the context of the excursion-set method. 
Among the three window functions, the real-space top-hat window function in Eq.~(\ref{eq:WF3}) 
requires careful treatment, as we will discuss in Section~\ref{subsec:ncov}. 
Although it is sometimes used in estimating the abundance of PBHs, its Fourier-space profile decays only slowly at large $z$, as can be seen in Figure~\ref{fig:wf}. 
This slow large-$z$ decay makes the variance of the noise sensitive to high-$z$ modes and leads to a qualitatively different small-$\widetilde\tau$ behavior from the other two window functions. 
Because of this difference, we will separately discuss the correlation properties of the noise for each window function below. 

\subsection{Noise covariance}
\label{subsec:ncov}

We here derive the covariances and the variances that will be implemented in our numerical simulations in Sections~\ref{stochastic} and~\ref{massfunction}. 
The formulas are prepared for each window function, whereas the tilt $\alpha$ of the power spectrum (\ref{eq:ps_alpha}) is fixed to be $\alpha = 0$ hereafter for simplicity, following Ref.~\cite{Kushwaha:2025zpz}. 

\paragraph*{\textit{Fourier-space top-hat window function.}} 

Let us start with the discontinuous window function, namely the Fourier-space top-hat window function (\ref{eq:WF1}). 
The quantities below have already appeared in Ref.~\cite{Kushwaha:2025zpz}, and we merely recap here for our later convenience. 
The covariance matrix (\ref{eq:app_xi_cov_general}) is given by 
\begin{align}
    \Delta^{2} 
    \expval*{ 
        \xi (\tau_{1}) \xi (\tau_{2}) 
    }_{\Theta} 
    &= 
    \frac{16}{81} 
    \frac{ \mathcal{P}_{0} \widetilde{k}_{\star} }{ \sqrt{2 \pi} } 
    \left( 
        \frac{4}{\widetilde{\tau}_{1}^{3} \widetilde{\tau}_{2}^{3}} 
        \left\{ 
            ( \widetilde{k}_{\star}^{2} + 2 ) 
            \exp \bigg( - \frac{ \widetilde{k}_{\star}^{2} }{2} \bigg) 
            - ( \widetilde{k}_{\star}^{2} + \widetilde{\tau}_{\wedge} \widetilde{k}_{\star} + \widetilde{\tau}_{\wedge}^{2} + 2 ) 
            \exp \bigg[ 
                - \frac{ ( \widetilde{\tau}_{\wedge} - \widetilde{k}_{\star} )^{2} }{2} 
            \bigg] 
        \right. 
    \right. 
    \notag \\ 
    &\quad \qquad\qquad\qquad\quad~ 
    \left. 
        + \, 
        \widetilde{k}_{\star} ( \widetilde{k}_{\star}^{2} + 3 ) 
        \sqrt{\frac{\pi}{2}} \, 
        \bigg[ 
            \mathrm{erf} 
            \bigg( 
                \frac{ \widetilde{k}_{\star} }{ \sqrt{2} } 
            \bigg) 
            + 
            \mathrm{erf} 
            \bigg( 
                \frac{ \widetilde{\tau}_{\wedge} - \widetilde{k}_{\star} }{ \sqrt{2} } 
            \bigg) 
        \bigg] 
    \right\} 
    \notag \\ 
    &\quad \qquad\qquad\qquad\quad~ 
    \left. 
        - \, 
        \frac{2 \widetilde{\tau}_{\wedge}}{ \widetilde{\tau}_{\vee}^{3}} 
        \exp \bigg[ 
            - \frac{ ( \widetilde{\tau}_{\wedge} - \widetilde{k}_{\star} )^{2} }{2} 
        \bigg] 
        + 
        \frac{ \delta_{\rm D} ( \widetilde{\tau}_{1} - \widetilde{\tau}_{2} ) }{ \widetilde{\tau}_{1} } 
        \exp \bigg[ 
            - \frac{ ( \widetilde{\tau}_{\wedge} - \widetilde{k}_{\star} )^{2} }{2} 
        \bigg] 
    \right) 
    \,\, ,
    \label{eq:ncov_W1}
\end{align}
where we have introduced 
\begin{equation*}
	\widetilde{\tau}_{\wedge} 
    \equiv \mathrm{min} (\widetilde{\tau}_{1}, \, \widetilde{\tau}_{2}) 
	\,\, , 
	\qquad 
	\widetilde{\tau}_{\vee} 
    \equiv \mathrm{max} (\widetilde{\tau}_{1}, \, \widetilde{\tau}_{2}) 
	\,\, . 
\end{equation*}
Deriving Eq.~(\ref{eq:ncov_W1}) from Eq.~(\ref{eq:app_xi_cov_general}), the integrals involving the product of two $\dd\, [ z^2\Theta(1-z) ] / \dd z = 2z\Theta(1-z) - z^2\delta_{\rm D}(1-z)$ are performed. 
The $\Theta \Theta$-term gives rise to the terms in the curly bracket in Eq.~(\ref{eq:ncov_W1}), and the cross terms to the one proportional to $2 \widetilde{\tau}_{\wedge} / \widetilde{\tau}_{\vee}^{3}$. 
The $\delta_{\rm D} \delta_{\rm D}$-term to the last term containing the $\delta$-function. 
This can be contrasted to the standard excursion-set method in a non-PBH context, where, with $\widetilde{W}_{\Theta} (z)$, only the last term in Eq.~(\ref{eq:ncov_W1}) is present. 
When the method is applied to the context of PBHs, however, the condition that a PBH forms when the relevant mode crosses the horizon gives the additional $z^{2}$ factor multiplied to the window function, which is differentiated further and results in the correlated parts. 

When we discretize the time variable $\widetilde{\tau}$ in our numerical simulations, the $\delta$-function in Eq.~(\ref{eq:ncov_W1}) is replaced by $\delta_{\rm D} (\widetilde{\tau}_{i} - \widetilde{\tau}_{j}) \to \delta_{ij} / \Delta \widetilde{\tau}$, where $\delta_{ij}$ is the Kronecker-$\delta$, and $\Delta \widetilde{\tau}$ is the mesh size. 
This also applies when one considers the limit $\widetilde{\tau}_{1} =\widetilde{\tau}_{2} = \widetilde{\tau}$, resulting in the variance of the noise, 
\begin{align}
    \expval*{ 
        [ \Delta \xi (\tau) ]^{2} 
    }_{\Theta} 
    &= 
    \frac{16}{81} 
    \frac{ \mathcal{P}_{0} \widetilde{k}_{\star} }{ \sqrt{2 \pi} } 
    \left( 
        \frac{4}{\widetilde{\tau}^{6}} 
        \left\{ 
            ( \widetilde{k}_{\star}^{2} + 2 ) 
            \exp \bigg( - \frac{ \widetilde{k}_{\star}^{2} }{2} \bigg) 
            - ( \widetilde{k}_{\star}^{2} + \widetilde{\tau} \widetilde{k}_{\star} + \widetilde{\tau}^{2} + 2 ) 
            \exp \bigg[ 
                - \frac{ ( \widetilde{\tau} - \widetilde{k}_{\star} )^{2} }{2} 
            \bigg] 
        \right. 
    \right. 
    \notag \\ 
    &\quad \qquad\qquad\qquad\quad~ 
    \left. 
        + \, 
        \widetilde{k}_{\star} ( \widetilde{k}_{\star}^{2} + 3 ) 
        \sqrt{\frac{\pi}{2}} \, 
        \bigg[ 
            \mathrm{erf} 
            \bigg( 
                \frac{ \widetilde{k}_{\star} }{ \sqrt{2} } 
            \bigg) 
            + 
            \mathrm{erf} 
            \bigg( 
                \frac{ \widetilde{\tau} - \widetilde{k}_{\star} }{ \sqrt{2} } 
            \bigg) 
        \bigg] 
    \right\} 
    \notag \\ 
    &\quad \qquad\qquad\qquad\quad~ 
    \left. 
        - \, 
        \frac{2}{ \widetilde{\tau}^2 } 
        \exp \bigg[ 
            - \frac{ ( \widetilde{\tau} - \widetilde{k}_{\star} )^{2} }{2} 
        \bigg] 
        + 
        \frac{ 1 }{ \widetilde{\tau} \Delta \widetilde{\tau} } 
        \exp \bigg[ 
            - \frac{ ( \widetilde{\tau} - \widetilde{k}_{\star} )^{2} }{2} 
        \bigg] 
    \right) 
    \,\, . 
    \label{eq:nvar_W1}
\end{align}
When $\widetilde{\tau} \ll 1$, the variance of the noise approaches 
\begin{equation}
    \expval*{ 
        [ \Delta \xi (\tau) ]^{2} 
    }_{\Theta} 
    \approx 
    \frac{16}{81} 
    \frac{ \mathcal{P}_{0} \widetilde{k}_{\star} }{ \sqrt{2 \pi} } 
    \frac{ \mathrm{exp} ( - \widetilde{k}_{\star}^{2} / 2 ) }{ \widetilde{\tau}^{2}} 
    \qty( \frac{\widetilde{\tau}}{ \Delta \widetilde{\tau} } - 1 ) 
    \,\, . 
    \label{eq:nvar_W1_smt}
\end{equation}
This implies that, though the variance diverges as $\widetilde{\tau} \to 0$ for a fixed $\widetilde{k}_{\star}$, one can implement numerical simulations with a small but nonzero initial $\widetilde{\tau}$ such that $\widetilde{\tau} \geq \Delta \widetilde{\tau}$. 
By doing so, the variance of the noise at the initial ``time'' is suppressed by $\mathrm{exp} (- \widetilde{k}_{\star}^{2} / 2)$ for $\widetilde{k}_{\star} \gtrsim \mathcal{O} (1)$, which is the case of our interest. 

The above covariance and variance, Eqs.~(\ref{eq:ncov_W1}) and (\ref{eq:nvar_W1}), determine the stochastic trajectories that will be numerically generated in Section~\ref{stochastic}. 
The generated trajectories are averaged over many realizations, and the consistency of our numerical simulations can be confirmed through Eq.~(\ref{eq:nvar_W1}) together with the variance of the density contrast, 
\begin{align}
    \expval*{ 
        [ \delta (\tau) ]^{2} 
    }_{\Theta} 
    &= \frac{16}{81} 
    \frac{ \mathcal{P}_{0} \widetilde{k}_{\star} }{ \sqrt{2 \pi} } 
    \, 
    \frac{1}{\widetilde{\tau}^{4}} 
    \left\{ 
        ( \widetilde{k}_{\star}^{2} + 2 ) \exp \bigg( 
            - \frac{\widetilde{k}_{\star}^{2}}{2} 
        \bigg) 
        - ( \widetilde{\tau}^{2} + \widetilde{k}_{\star} \widetilde{\tau} + \widetilde{k}_{\star}^{2} + 2 ) 
        \exp \bigg[ 
            - \frac{( \widetilde{\tau} - \widetilde{k}_{\star} )^{2}}{2} 
        \bigg] 
    \right. 
    \notag \\ 
    &\quad \qquad\qquad\qquad 
    \left. 
        + \, 
        \widetilde{k}_{\star} ( \widetilde{k}_{\star}^{2} + 3 ) \sqrt{\frac{\pi}{2}} \, 
        \qty[ 
            \mathrm{erf} \bigg( 
                \frac{\widetilde{k}_{\star}}{\sqrt{2}} 
            \bigg) 
            + 
            \mathrm{erf} \bigg( 
                \frac{\widetilde{\tau} - \widetilde{k}_{\star}}{\sqrt{2}} 
            \bigg) 
        ] 
    \right\} 
    \,\, . 
    \label{eq:dvar_W1}
\end{align}

\paragraph*{\textit{Gaussian window function.}} 

We now turn to the Gaussian window function (\ref{eq:WF2}), which is a smooth function and still enables us to derive the following statistical quantities in closed form. 
For $\alpha=0$, the covariance of the noise is given by 
\begin{align}
    \Delta^{2} \expval*{ 
            \xi (\tau_{1}) \xi (\tau_{2}) 
    }_{\rm G} 
    &= 
    \frac{16}{81} 
    \frac{ \mathcal{P}_{0} \widetilde{k}_{\star} }{ \sqrt{2 \pi} } 
    \, 
    \exp \bigg( - \frac{ \widetilde{k}_{\star}^{2} }{2} \bigg) 
    \frac{ \widetilde{\tau}_{1}^{9} \widetilde{\tau}_{2}^{9} }{ ( \widetilde{\tau}_{1}^{2} + \widetilde{\tau}_{1}^{2} \widetilde{\tau}_{2}^{2} + \widetilde{\tau}_{2}^{2} )^{7} } 
    \bigg( 
        ( 
            \widetilde{k}_{\star}^{6} 
            + 20 \widetilde{k}_{\star}^{4} f^{2} 
            + 87 \widetilde{k}_{\star}^{2} f^{4} 
            + 48 f^{6} 
        ) 
    \bigg. 
    \notag \\ 
    &\quad 
    + \, 
    2 \widetilde{\tau}_{1}^{2} \widetilde{\tau}_{2}^{2} f^{4} 
    [ 
        (1 - f^{2}) \widetilde{k}_{\star}^{4} 
        + f^{2} (9 - 7f^{2}) \widetilde{k}_{\star}^{2} 
        + 4 f^{4} (2 - f^{2}) 
    ] 
    \notag \\ 
    &\quad 
    + \sqrt{\pi} \, 
    \bigg( 
        \frac{ \widetilde{k}_{\star} }{ \sqrt{2} \, f }
    \bigg) 
    \exp \bigg( 
        \frac{ \widetilde{k}_{\star}^{2} }{ 2 f^{2} } 
    \bigg) 
    \bigg[ 
        1 + \mathrm{erf} \bigg( \frac{ \widetilde{k}_{\star} }{ \sqrt{2} \, f } \bigg) 
    \bigg] 
    \bigg\{ 
        ( 
            \widetilde{k}_{\star}^{6} 
            + 21 \widetilde{k}_{\star}^{4} f^{2} 
            + 105 \widetilde{k}_{\star}^{2} f^{4} 
            + 105 f^{6} 
        ) 
    \bigg. 
    \notag \\ 
    &\quad \qquad
    \bigg. 
        \bigg. 
            + 
            \, 
            2 \widetilde{\tau}_{1}^{2} \widetilde{\tau}_{2}^{2} f^{4} 
            [ 
                (1 - f^{2}) \widetilde{k}_{\star}^{4} 
                + 2 f^{2} (5 - 4 f^{2}) \widetilde{k}_{\star}^{2} 
                + 3 f^{4} (5 - 3 f^{2}) 
            ] 
        \bigg\} 
    \bigg) 
    \,\, . 
    \label{eq:ncov_W2}
\end{align}
In Eq.~(\ref{eq:ncov_W2}), we have introduced the function 
\begin{equation}
    f 
    = f (\widetilde{\tau}_{1}, \, \widetilde{\tau}_{2}) 
    \equiv 
    \sqrt{ 1 + \frac{1}{\widetilde{\tau}_{1}^{2}} + \frac{1}{\widetilde{\tau}_{2}^{2}} } 
    \,\, . 
\end{equation}
The equal-time limit can now be taken straightforwardly, just by setting $\widetilde{\tau}_{1} = \widetilde{\tau}_{2} = \widetilde{\tau}$ in Eq.~(\ref{eq:ncov_W2}). 
For $\widetilde{\tau} \ll 1$, it asymptotes to 
\begin{equation}
    \Delta^{2} \expval*{ 
            \xi (\tau_{1}) \xi (\tau_{2}) 
    }_{\rm G} 
    \approx  
    \frac{16}{81} 
    \frac{ \mathcal{P}_{0} \widetilde{k}_{\star} }{ \sqrt{2 \pi} } 
    \frac{ \mathrm{exp} ( - \widetilde{k}_{\star}^{2} / 2 ) }{ \widetilde{\tau}^{2}} 
    \,\, . 
    \label{eq:nvar_W2_smt}
\end{equation}
As in Eq.~(\ref{eq:nvar_W1_smt}), Eq.~(\ref{eq:nvar_W2_smt}) also remains finite as long as the initial $\widetilde{\tau}$ is chosen to be small enough but nonzero. 

For consistency check of our numerical simulations, the variance of the coarse-grained density contrast $\delta (\tau)$ is used, which is given by 
\begin{align}
    \expval*{ [ \delta (\tau) ]^{2} }_{\rm G} 
    &= \frac{16}{81} 
    \frac{\mathcal{P}_{0} \widetilde{k}_{\star}}{\sqrt{2 \pi}} \, 
    \frac{ \widetilde{\tau}^{2} }{ ( \widetilde{\tau}^{2} + 2 )^{3} } 
    \exp \bigg( - \frac{ \widetilde{k}_{\star}^{2} }{2} \bigg) 
    \left\{ 
        \vphantom{
            \qty[ 
                1 + \mathrm{erf} \bigg( \frac{ \widetilde{k}_{\star}}{ \sqrt{2} \, f} \bigg) 
            ] 
        } 
        ( \widetilde{k}_{\star}^{2} + 2 f^{2} ) 
    \right. 
    \notag \\ 
    &\quad 
    \left. 
        + \, \sqrt{\pi} \, 
        \bigg( \frac{ \widetilde{k}_{\star} }{ \sqrt{2} \, f } \bigg) 
        \exp \bigg( \frac{ \widetilde{k}_{\star}^{2} }{ 2 f^{2} } \bigg) 
        \qty[ 
            1 + \mathrm{erf} \bigg( \frac{ \widetilde{k}_{\star}}{ \sqrt{2} \, f} \bigg) 
        ] 
        ( \widetilde{k}_{\star}^{2} + 3 f^{2} ) 
    \right\} 
    \,\, . 
    \label{eq:dvar_W2}
\end{align}

\paragraph*{\textit{Real-space top-hat window function.}} 

Now, let us turn our attention to the top-hat window function in real-space, Eq.~(\ref{eq:WF3}). 
For the window function, the variance of the noise follows from Eq.~(\ref{eq:xi_variance_z}) that 
\begin{equation}
    \expval*{ [ \Delta \xi (\tau) ]^{2} }_{\rm TH} 
    = \frac{16}{81} 
    \frac{\mathcal{P}_{0} \widetilde{k}_{\star}}{\sqrt{2 \pi}} \, 
    \frac{9}{\widetilde{\tau}^{2}} 
    \int_{0}^{\infty} 
    \dd z \, 
    z  
    \exp \bigg[ 
        - \frac{\widetilde{\tau}^{2}}{2} 
        \bigg( 
            z - \frac{ \widetilde{k}_{\star} }{ \widetilde{\tau} } 
        \bigg)^{2}  
    \bigg] 
    [ 
        z j_{0} (z) - j_{1} (z) 
    ]^{2} 
    \,\, . 
    \label{eq:var_W3is}
\end{equation}
While the integral in Eq.~(\ref{eq:var_W3is}) does not converge for $\widetilde{\tau} \to 0$, let us observe the asymptotic behavior in the $\widetilde{\tau} \ll 1$ limit. 
The dominating saddle of the integral resides at $z = \widetilde{k}_{\star} / \widetilde{\tau}$, which becomes larger as $\widetilde{\tau}$ approaches zero. 
Therefore, we approximate the last factor as $z j_{0} (z) - j_{1} (z) \approx \sin z$, having a large $z$ in mind. 
With this approximation, it is more convenient to rewrite the integral in terms of $x = \widetilde{\tau} z$, as 
\begin{equation}
    \expval*{ [ \Delta \xi (\tau) ]^{2} }_{\rm TH} 
    \approx \frac{16}{81} 
    \frac{\mathcal{P}_{0} \widetilde{k}_{\star}}{\sqrt{2 \pi}} \, 
    \frac{9}{\widetilde{\tau}^{4}} 
    \int_{0}^{\infty} 
    \dd x \, x 
    \exp \bigg[ 
        - \frac{ (x - \widetilde{k}_{\star})^{2} }{2} 
    \bigg] 
    \sin^{2} \bigg( \frac{x}{\widetilde{\tau}} \bigg) 
    \,\, . 
\end{equation}
The last factor $\sin^{2} (x / \widetilde{\tau})$ shows a highly oscillatory behavior for small $\widetilde{\tau}$, thereby replacing it with its mean, $1/2$. 
One then arrives at 
\begin{equation}
    \expval*{ [ \Delta \xi (\tau) ]^{2} }_{\rm TH} 
    \approx \frac{16}{81} 
    \frac{\mathcal{P}_{0} \widetilde{k}_{\star}}{\sqrt{2 \pi}} \, 
    \frac{1}{\widetilde{\tau}^{4}} 
    \times \frac{9}{2} 
    \bigg\{ 
        \exp \bigg( 
            - \frac{\widetilde{k}_{\star}^{2}}{2} 
        \bigg) 
        + \widetilde{k}_{\star} 
        \sqrt{\frac{\pi}{2}} 
        \, 
        \bigg[ 
            1 + \mathrm{erf} \bigg( 
                \frac{\widetilde{k}_{\star}}{\sqrt{2}} 
            \bigg) 
        \bigg] 
    \bigg\} 
    \,\, . 
    \label{eq:var_W3is_asp}
\end{equation}
A difference from the cases with the previous two window functions can be observed in Eq.~(\ref{eq:var_W3is_asp}). 
That is, for $\widetilde{k}_{\star} \gtrsim \mathcal{O} (1)$, the dominant contribution in the curly bracket is $\widetilde{k}_{\star} \cdot \sqrt{\pi / 2} \cdot 2 = \sqrt{2 \pi} \, \widetilde{k}_{\star}$, instead of the exponentially suppressing factor $\mathrm{exp} ( - \widetilde{k}_{\star}^{2} / 2)$. 
This means that, even if one starts from an initial and small but nonzero $\widetilde{\tau}$, the noise is too large so that it would not result in a physically reasonable simulation. 
For this reason, following Refs.~\cite{Ando:2018qdb, Young:2019osy, Yoo:2020dkz, Young:2024jsu}, we will take the transfer function into account in what follows, whenever the real-space window function is used\footnote{%
    Indeed, it is known that the use of the transfer function may be neglected when either the Fourier-space top-hat or the Gaussian window function is used~\cite{Ando:2018qdb}.  
}. 

To proceed, for $\widetilde{W} (z) = \widetilde{W}_{\rm TH} (z)$, we replace $\widetilde{W} (z)$ in the previous equations with 
\begin{equation}
    \widetilde{W} (z) 
    \longrightarrow 
    \widetilde W_{\rm eff}(z)
    \equiv
    \widetilde W(z) T \qty( \frac{z}{\sqrt{3}}) 
    \,\, , 
    \label{eq:repw}
\end{equation}
where the function $T (z / \sqrt{3})$ is the radiation transfer function for $T(y) \equiv 3 j_1(y) / y$.
With Eq.~(\ref{eq:covariancexi}) and Eq.~(\ref{eq:repw}), the covariance matrix of the noise is given by 
\begin{align}
    \expval*{
        \xi (\tau_{1}) \xi (\tau_{2}) 
    }_{\rm eff} 
    &= \frac{16}{81} 
    \frac{1}{\tau_{1}^{2} \tau_{2}^{2}} 
    \int_{0}^{\infty} \frac{\dd k}{k} 
    \, \mathcal{P}_{\zeta} (k) 
    k^{2} 
    \notag \\ 
    &\quad \times 
    \qty{ 
        \dv{z_{1}} [ z_{1}^{2} \widetilde{W}_{\rm eff} (z_{1}) ]
    }_{z_{1} = k / \tau_{1}} 
    \qty{ 
        \dv{z_{2}} [ z_{2}^{2} \widetilde{W}_{\rm eff} (z_{2}) ]
    }_{z_{2} = k / \tau_{2}} 
    \notag \\  
&= \frac{48\Delta^{-2}}{\widetilde\tau_1^2\widetilde\tau_2^2}\frac{\mathcal P_0\widetilde k_\star}{\sqrt{2\pi}}\int_0^\infty
    \frac{\dd x}{x}x^{2}\exp\qty[-\frac{( x - \widetilde{k}_{\star} )^{2}}{2}] 
    \notag \\ 
    &\quad \times 
    \qty{ 
        \dv{z_{1}} \qty[ 
            j_{1} (z_{1}) 
            j_{1} \qty(\frac{z_{1}}{\sqrt{3}} ) 
        ]
    }_{z_{1} = x / \widetilde{\tau}_{1}} \qty{ 
        \dv{z_{2}} \qty[        
            j_{1} (z_{2}) 
            j_{1} \qty( \frac{z_{2}}{\sqrt{3}} ) 
        ]
    }_{z_{2} = x / \widetilde{\tau}_{2}} 
    \,\, . 
    \label{eq:ncov_W3tf}
\end{align}
The variance of the noise becomes 
\begin{align}
    \expval*{ [ \Delta \xi (\tau) ]^{2} }_{\rm eff}
    &= 
    \frac{16}{81}
    \frac{\mathcal{P}_{0}}{ \sqrt{2 \pi} } 
    \,
    \frac{ \widetilde{k}_{\star}}{ \widetilde{\tau}^{ 2}} 
    \int_{0}^{\infty} \dd z \, z^{ 3} 
    \exp \bigg[ 
        - \frac{\widetilde{\tau}^{2}}{2} 
        \bigg( z - \frac{ \widetilde{k}_{\star} }{ \widetilde{\tau} } \bigg)^{2} 
    \bigg] 
    \notag \\
    &\quad \times 
    81 
    \bigg[ 
        j_{0} (z) \frac{ j_{1} (z / \sqrt{3}) }{ z / \sqrt{3} } 
        + j_{0} \qty( \frac{z}{\sqrt{3}} ) \frac{j_{1} (z)}{z} 
        - 4 \frac{ j_{1} (z) }{z} \frac{ j_{1} (z / \sqrt{3}) }{ z / \sqrt{3} } 
    \bigg]^{2} 
    \,\, . 
    \label{eq:varince_noise_real}
\end{align}

The variance of the density contrast is
\begin{align}
    \expval*{ [ \delta (\tau) ]^{2} }_{\rm eff}
    &= 
    \frac{16}{81} 
    \frac{ \mathcal{P}_{0} \widetilde{k}_{\star} }{\sqrt{2 \pi}} 
    \int_{0}^{\infty} 
    \dd z \, 
    z^{3} 
    \exp \bigg[ - \frac{ \widetilde{\tau}^{2} }{2} \bigg( 
            z - \frac{ \widetilde{k}_{\star}}{\widetilde{\tau}} 
        \bigg)^{2} 
    \bigg] 
    \bigg[ 
        9 \frac{ j_{1} (z) }{z} \frac{ j_{1} (z / \sqrt{3}) }{ z / \sqrt{3} } 
    \bigg]^{2} 
    \,\, . 
    \label{eq:dvar_W3tf}
\end{align}
We will use Eq.~(\ref{eq:ncov_W3tf}) to numerically generate the noise with the desired covariance, instead of the counterpart without the transfer function, and the variances given by Eq.~(\ref{eq:varince_noise_real}) and Eq.~(\ref{eq:dvar_W3tf}) to check the consistency of our numerical simulations. 

\vspace{1.0\baselineskip}

Before closing this section, we give the small-$\widetilde{\tau}$ limit for a general $\alpha$, since the tilt $\alpha$ is related to the behavior at infrared (\textit{i.e.}~large-scale) limit of the variance of both noise and coarse-grained density contrast. 
For the three window functions, Eqs.~(\ref{eq:WF1}), (\ref{eq:WF2}), and (\ref{eq:WF3}), the variance of the noise  in the $\widetilde{\tau} \to 0$ limit respectively behaves 
\begin{align}
    \expval*{ 
        [\Delta \xi (\tau) ]^{2} 
    } 
    &= 
    \frac{16}{81} 
    \frac{ \mathcal{P}_{0} }{ \sqrt{2 \pi} } 
    \frac{1}{\widetilde{k}_{\star}^{\alpha - 1}} 
    \notag \\[2.0ex] 
    &\quad \times 
    \begin{cases}
        \displaystyle ~ 
        \frac{\widetilde{\tau}^{\alpha}}{ \widetilde{\tau}^{2} } 
        \exp \bigg( - 
            \frac{\widetilde{k}_{\star}^{2}}{2} 
        \bigg) 
        \qty[ 
            \frac{\widetilde{\tau}}{\Delta \widetilde{\tau}} 
            - \frac{2\qty(\alpha+2)}{\alpha+4} 
        ] 
        &\quad \text{for} \quad \widetilde{W}_{\Theta} (z) 
        \,\, , 
        \\[2.5ex] 
        \displaystyle ~ 
        \frac{ \widetilde{\tau}^{\alpha} }{ \widetilde{\tau}^{2} } 
        \exp \bigg( 
            - \frac{\widetilde{k}_{\star}^{2}}{2} 
        \bigg) 
        \frac{\alpha^{2} + 2 \alpha + 8}{8} 
        \Gamma \qty( \frac{\alpha + 4}{2} ) 
        &\quad \text{for} \quad \widetilde{W}_{\rm G} (z) 
        \,\, , 
        \\[2.5ex] 
        \displaystyle ~ 
        \frac{F_{\alpha} (\widetilde{k}_{\star})}{\widetilde{\tau}^{4}} 
        &\quad \text{for} \quad \widetilde{W}_{\rm TH} (z) 
        \,\, . 
        \end{cases}
        \label{eq:varxi_asymp}
\end{align}
In Eq.~(\ref{eq:varxi_asymp}), the function $F_{\alpha} (\widetilde{k}_{\star})$ is defined through the confluent hypergeometric function by 
\begin{equation}
    \frac{2}{9} F_{\alpha} (\widetilde{k}_{\star}) 
    \equiv 
    2^{\alpha / 2} 
    \qty[ 
        \Gamma \qty( 1 + \frac{\alpha}{2} ) 
        {}_{1} F_{1} \bigg( 
            \begin{matrix} 
                - (\alpha + 1) / 2 
                \\ 
                1/2 
            \end{matrix} 
            ~\middle|~ 
            - \frac{\widetilde{k}_{\star}^{2}}{2} 
        \bigg) 
        + \sqrt{2} \, \widetilde{k}_{\star} 
        \Gamma \qty( \frac{\alpha + 3}{2} ) 
        {}_{1} F_{1} \bigg( 
            \begin{matrix} 
                - \alpha / 2 
                \\ 
                3/2 
            \end{matrix} 
            ~\middle|~ 
            - \frac{\widetilde{k}_{\star}^{2}}{2} 
        \bigg) 
    ] 
    \,\, . 
    \label{eq:fnf}
\end{equation}

In Eq.~(\ref{eq:varxi_asymp}), it can be seen that the tilt $\alpha \geq 0$ of the power spectrum determines the behavior of the variance of the noise in the vicinity of $\widetilde{\tau} = 0$. 
For the Fourier-space top-hat and the Gaussian window functions, an appropriate choice of $\alpha$ avoids the divergence of $\expval*{ [ \Delta \xi (\tau) ]^{2} }$ at $\widetilde{\tau} = 0$, which sometimes becomes an obstacle when implementing the noise numerically. 
Though a nonzero positive $\alpha$ may be chosen to alleviate or remove the divergence, it is not necessary to do so for the reason described below Eq.~(\ref{eq:nvar_W1_smt}). 
This, on the other hand, ceases to be the case when the real-space top-hat window function is used, in which it diverges in proportion to $1 / \widetilde{\tau}^{4}$ regardless of $\alpha$. 
Meanwhile, the transfer function makes $\widetilde{W}_{\rm eff} (z)$ more quickly decaying for large $z$, thereby keeping the behavior in the $\widetilde{\tau} \to 0$ limit finite as can be seen in Figure~\ref{fig:var}. 

\section{Stochastic trajectories and excursion-set probabilities}
\label{stochastic}

Having derived the statistical quantities of the coarse-grained density contrast, we now numerically solve the stochastic process of $\delta (\tau)$, with respect to the coarse-graining variable $\tau$. 
In any case, regardless of the choice of the window function, the Gaussianity of the coarse-grained density contrast is assumed. 
What is characteristic of our context is that the stochastic noises are correlated across scales, with the covariance determined by the primordial curvature power spectrum and the window function. 
Those involved factors, unfortunately, prohibit us from further proceeding analytically. 
We therefore generate stochastic trajectories numerically from this covariance matrix, using the scheme described in Section~\ref{sec:noise}. 
The resulting trajectories exhibit nontrivial correlations across coarse-graining scales, as we will see in Section~\ref{subsec:straj}.
From the generated trajectories, the formation probabilities for PBHs 
can be extracted, which will be discussed in Section~\ref{subsec:fprob}.

\subsection{Generation of colored noises}
\label{sec:noise}

To implement the statistical properties in our numerical simulations, we discretize the coarse-graining (``time'') variable $\widetilde{\tau} = \tau / \Delta$ on a finite grid. 
For the cases with smooth window functions, we use a mixed grid with $N_{\tau} = 800$ points. 
The grid starts from $\widetilde{\tau}_{\rm ini} = 10^{-5}$ and extends to $\widetilde{\tau}_{\rm max} = 10^{3}$. 
The interval $10^{-5}\leq \widetilde\tau<0.2$ is sampled logarithmically with $200$ points, $0.2\leq\widetilde\tau<2$ linearly with 400 points, and $2\leq\widetilde\tau\leq10^3$ logarithmically with $200$ points. The first grid spacing $\Delta\tau_{\rm ini}$ is
\begin{equation}
    \Delta \widetilde{\tau}_{\rm ini} 
    \simeq 
    5 \times 10^{-7} 
    \,\, , 
    \qquad 
    \frac{\widetilde{\tau}_{\rm ini}}{\Delta \widetilde{\tau}_{\rm ini}} 
    \simeq 
    20 
    \,\, .
\end{equation}
For the Fourier-space top-hat case, we instead use a uniform grid $\widetilde{\tau}_{i} = 0.01 \,i$, corresponding to $\widetilde{\tau}_{\rm ini} = 0.01$, $\Delta \widetilde{\tau} = 0.01$, and $\widetilde{\tau}_{\rm ini} / \Delta \widetilde{\tau} = 1$. 
On this grid, we construct the covariance matrix from either Eq.~(\ref{eq:ncov_W1}), (\ref{eq:ncov_W2}), or (\ref{eq:ncov_W3tf}), as 
\begin{equation}
    \Sigma_{ij} 
    \equiv 
    \Delta^{2} \expval*{ \xi (\tau_{i}) \xi (\tau_{j}) } 
    \,\, , 
    \label{eq:ch_cov}
\end{equation}
where the discretized and numerical covariance matrix $\Sigma_{ij}$ is evaluated from the covariance matrix analytically derived in Section~\ref{subsec:ncov}. 
Since the primordial curvature perturbation is assumed to be Gaussian and its linear relation to the coarse-grained density contrast is assumed, the noise vector is distributed as a multivariate Gaussian, $\boldsymbol{\xi} \sim \mathcal N (\boldsymbol{0}, \, \Sigma)$, where 
$\boldsymbol{\xi} = ( \xi (\tau_{1}), \, \xi (\tau_{2}), \, \dots, \, \xi (\tau_{N_{\tau}}) )^{\rm T}$, the mean vector is $\boldsymbol{0} = (0, \, 0, \, \cdots, \, 0)^{\rm T}$, and the covariance matrix is given by Eq.~(\ref{eq:ch_cov}). 

In general, numerical generation of the noise itself under the desired correlation properties is nontrivial. 
Meanwhile, the Cholesky factorization method has been used in several literature~\cite{Nikakhtar:2018qqg, Kushwaha:2025zpz, Saito:2025sny}, which we also follow here. 
In this scheme, the covariance matrix is factorized into the lower-triangular matrix $L$ and its transpose as 
\begin{equation}
    \Sigma = L L^{\rm T} 
    \,\, ,
\end{equation}
with $L$ also called the Cholesky factor. 
With $x_j$ being independent standardized Gaussian random variables satisfying $\left\langle x_i x_j \right\rangle = \delta_{ij}$, then one realization of the correlated stochastic noise is obtained as
\begin{equation}
    \Delta\xi(\tau_i) = \sum_j L_{ij} x_j .
\end{equation}
By construction, this satisfies
\begin{equation}
    \Delta^{2} 
    \expval*{ \xi (\tau_{i}) \xi (\tau_{j}) } 
    = 
    \sum_{k, \, \ell} L_{ik} L_{j \ell} \expval*{ x_{k} x_{\ell} } 
    = 
    \sum_{k, \, \ell} L_{ik} \delta_{k \ell} L_{j \ell} 
    = 
    \sum_{k} L_{ik} L_{jk} 
    = 
    L L^{\rm T} 
    = 
    \Sigma_{ij} 
    \,\, .
\end{equation}
The corresponding trajectory of the coarse-grained density contrast is then obtained by numerically integrating the noise,
\begin{equation}
    \delta(\widetilde{\tau}_{i}) 
    = 
    \int^{\widetilde{\tau}_{i}}_{0} 
    \dd \widetilde{\tau} \, 
    \Delta \xi(\widetilde{\tau}) 
    \,\, .
\end{equation}
In practice, this integral is evaluated on the discretized $\tau$-grid.
Repeating this procedure, drawing a new independent Gaussian vector, namely $x_{j}$, constructing the correlated noise $\xi (\tau_{i})$, and then integrating it to obtain $\delta (\tau_{i})$, yields the ensemble of stochastic trajectories of $\delta (\widetilde{\tau})$ used in our numerical calculations. 

\subsection{Stochastic trajectories}
\label{subsec:straj}

The stochastic trajectories in our excursion-set analyses provide the formation probabilities defined in Section~\ref{subsec:fprob} as well as the mass function of PBHs discussed in Section~\ref{massfunction}. 
For each realization and for each scale $\tau$, we ask whether or not the coarse-grained density contrast exceeds a given threshold, 
\begin{equation}
    \delta (\tau) 
    > \delta_{\rm c} 
    \,\, .
\end{equation}
The numerical results in the following use the parameters
\begin{gather}
    \widetilde{k}_{\star} = 8.0 
    \,\, , 
    \qquad 
    \delta_{\rm c} = 0.45 
    \,\, .
    \label{eq:params}
\end{gather}
The latter is a fiducial threshold for PBH formation in the radiation-dominated era \cite{Musco:2004ak, Harada:2013epa, Musco:2018rwt}. 

\begin{figure}
  \centering
  \begin{minipage}{0.46\linewidth}
    \centering
    \includegraphics[width=\linewidth]{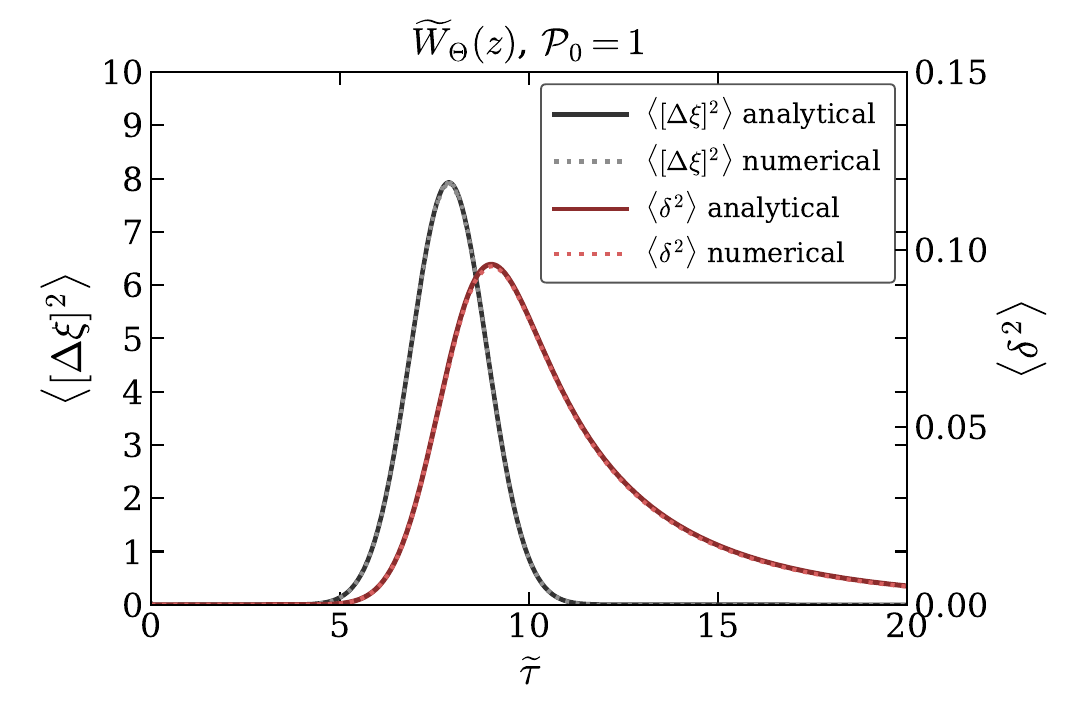}
    \label{fig:fb1}
  \end{minipage}
  \hspace{5mm}
  \begin{minipage}{0.46\linewidth}
    \centering
    \includegraphics[width=\linewidth]{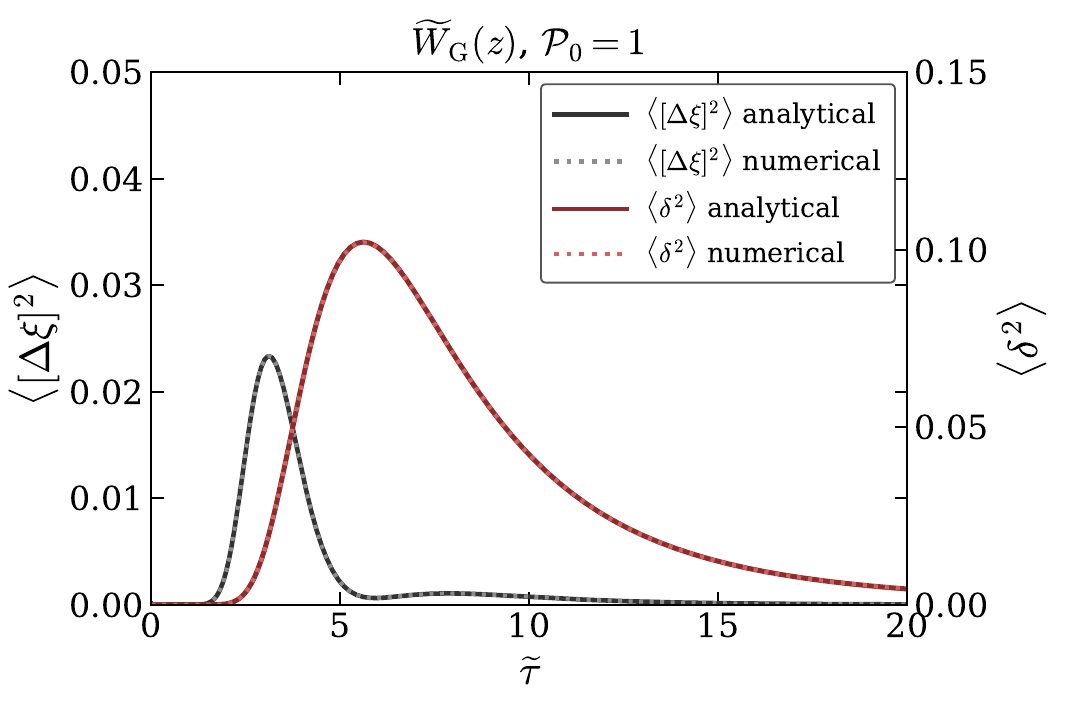}
    \label{fig:fb2}
  \end{minipage}
  \vspace{2.5mm}
  \begin{minipage}{0.46\linewidth}
    \centering
    \includegraphics[width=\linewidth]{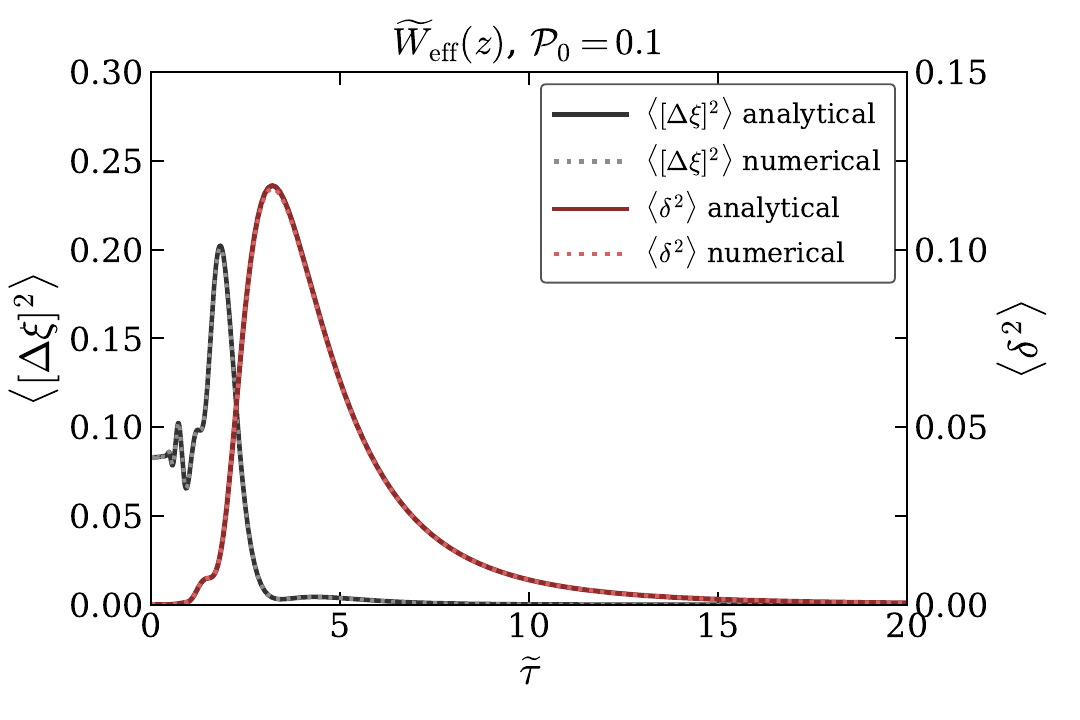}
    \label{fig:fb3-1eM1}
  \end{minipage}
  \hspace{5mm}
  \begin{minipage}{0.46\linewidth}
    \centering
    \includegraphics[width=\linewidth]{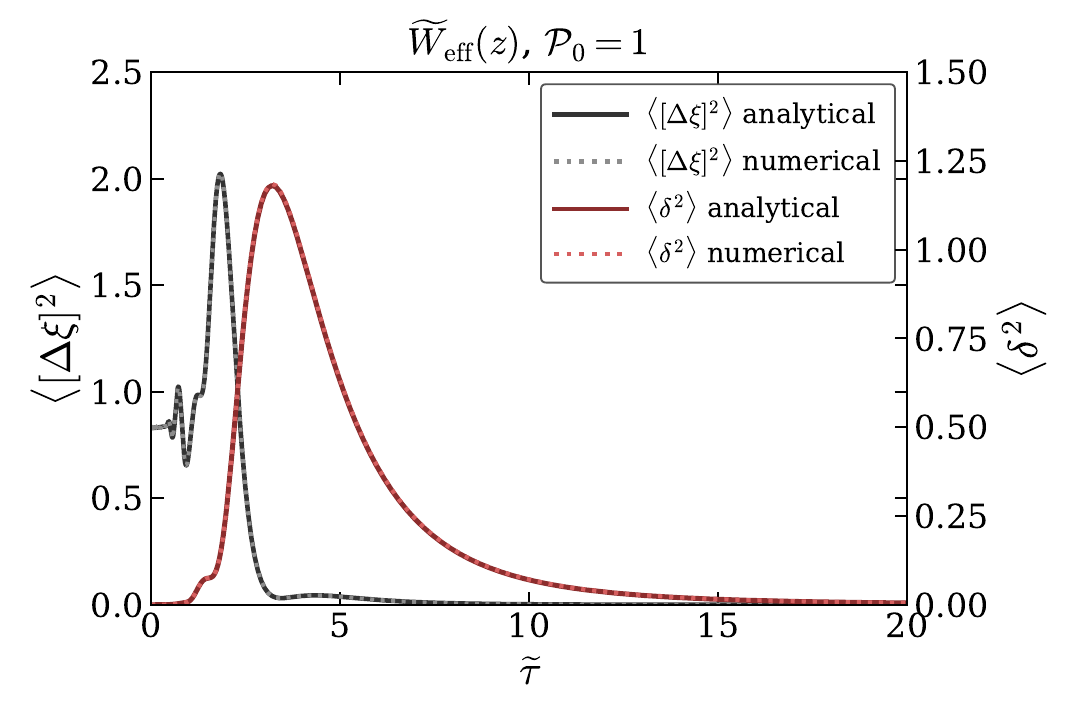}
    \label{fig:fb3-1e0}
  \end{minipage}
  \caption{
    Variances of the density contrast $\delta$ and the noise $\xi$ for the three kinds of window functions, Fourier-space top-hat (\textit{top left}), Gaussian (\textit{top right}), real-space top-hat with the transfer function (\textit{bottom}). 
    The amplitudes of the power spectrum, $\mathcal{P}_{0}$, are different between the bottom two panels. 
  }
  \label{fig:var}
\end{figure}

Figure~\ref{fig:var} shows the variances of the density contrast $\delta (\tau)$ and the stochastic noise $\xi (\tau)$ obtained from the samples of the numerical trajectories generated, which agree with the analytic formula given in Section~\ref{subsec:ncov} and thus support the consistency of our numerical simulations. 
The two bottom panels in Figure~\ref{fig:var} further show that, in the real-space top-hat window function with the transfer function case, the transfer function suppresses the strong small-$\widetilde{\tau}$ divergence of the variance of the noise associated with the slow large-$z$ decay of the real-space top-hat window function. 

\begin{figure}
  \centering
  \begin{minipage}{0.46\linewidth}
    \centering
    \includegraphics[width=\linewidth]{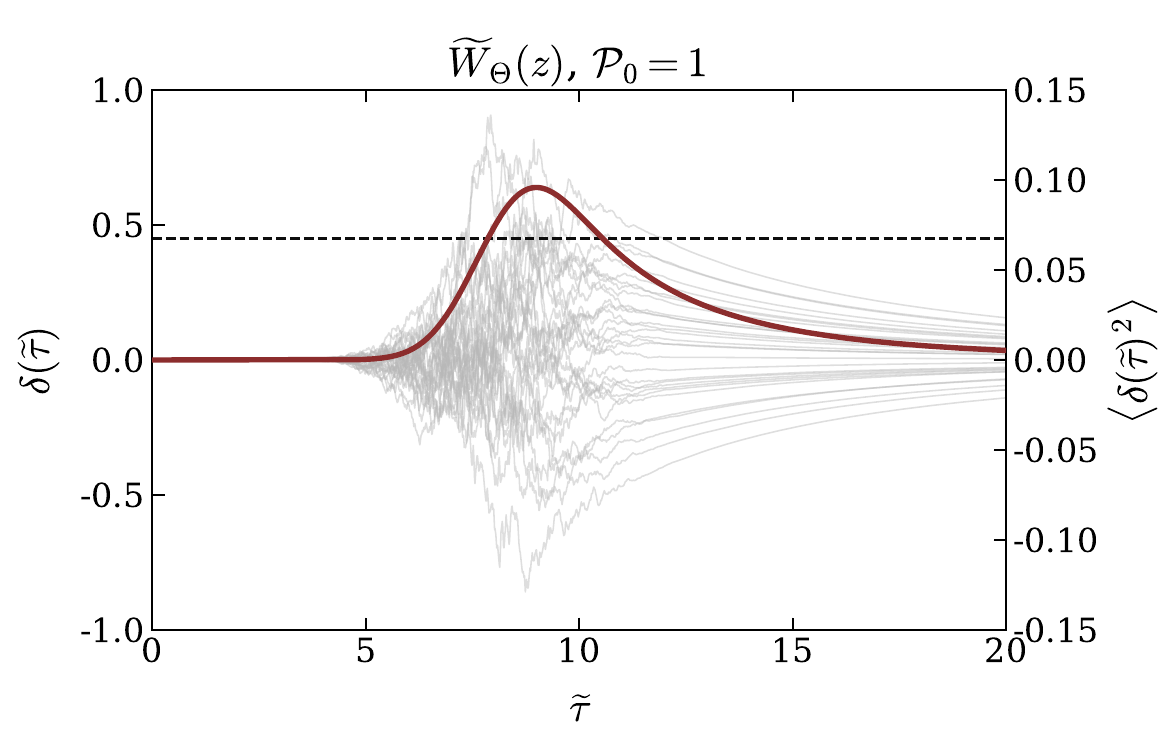}
    \label{fig:rw1}
  \end{minipage}
  \hspace{5mm}
  \begin{minipage}{0.46\linewidth}
    \centering
    \includegraphics[width=\linewidth]{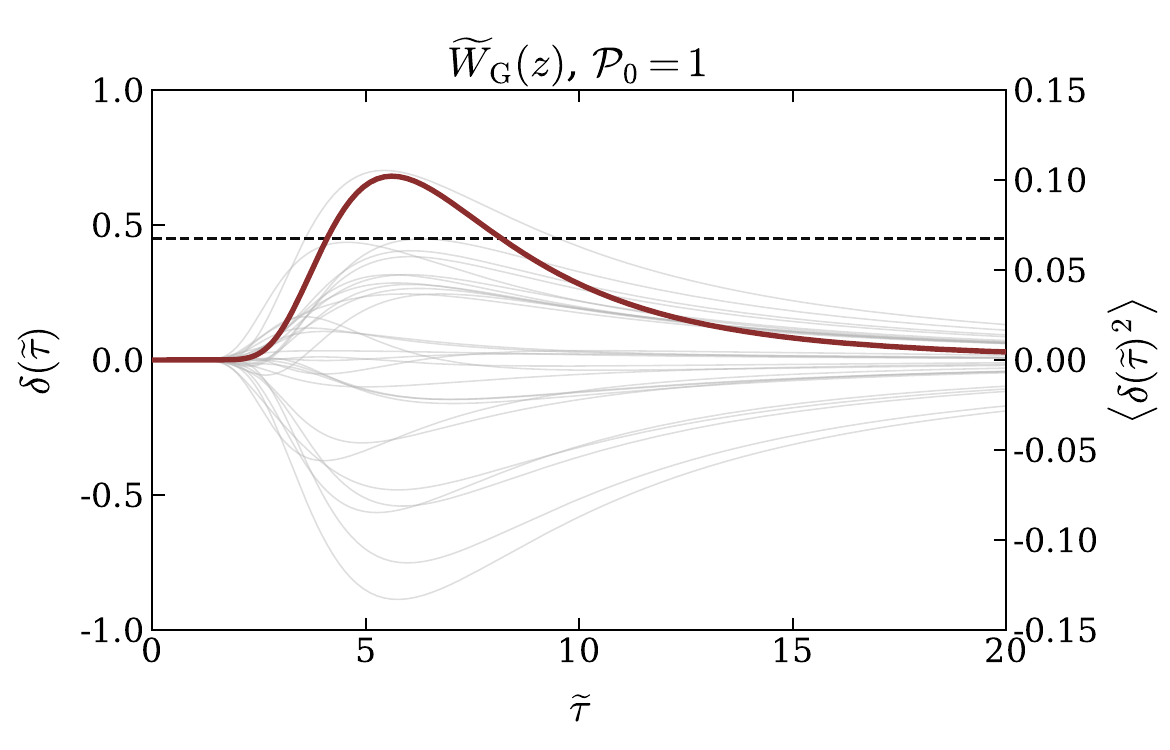}
    \label{fig:rw2}
  \end{minipage}
  \vspace{2.5mm}
  \begin{minipage}{0.46\linewidth}
    \centering
    \includegraphics[width=\linewidth]{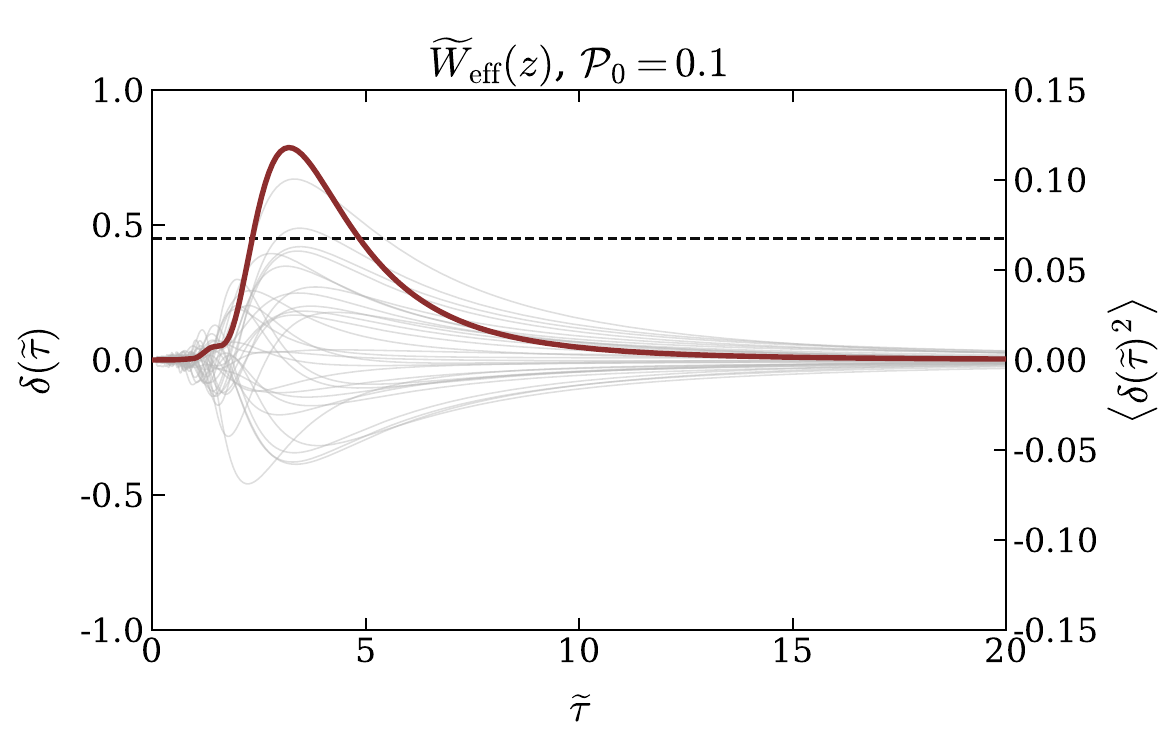}
    \label{fig:rw3-1eM1}
  \end{minipage}
  \hspace{5mm}
  \begin{minipage}{0.46\linewidth}
    \centering
    \includegraphics[width=\linewidth]{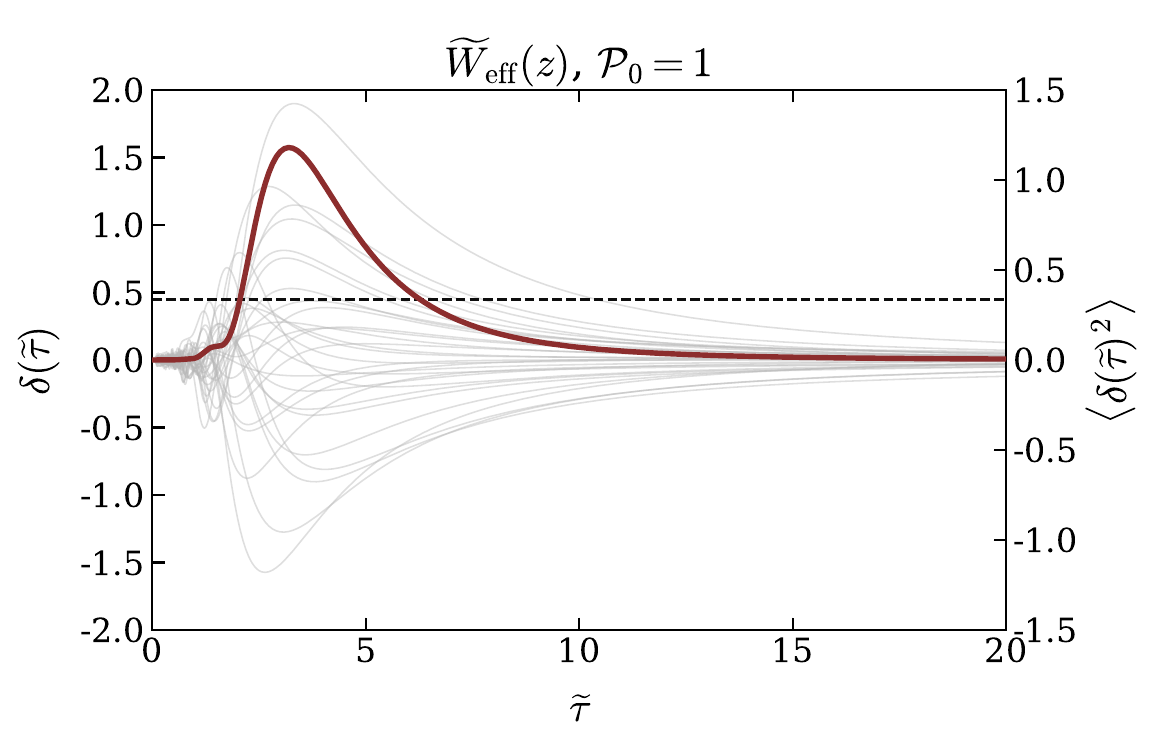}
    \label{fig:rw3-1e0}
  \end{minipage}
  \caption{
    Twenty sample stochastic trajectories of the density contrast, coarse-grained by the three kinds of window functions, Fourier-space top-hat (\textit{top left}), Gaussian (\textit{top right}), and real-space top-hat (\textit{bottom}). 
    The amplitudes of the power spectrum, $\mathcal{P}_{0}$, are different between the bottom two panels. 
    The analytical variance of the coarse-grained density contrast $\expval*{ [ \delta (\tau) ]^{2} }$ is shown by the wine-red thick curve. 
  }
  \label{fig:straj}
\end{figure}

Figure~\ref{fig:straj} displays $20$ representative realizations of $\delta (\tau)$ generated by the scheme explained in Section~\ref{sec:noise}, out of the total $N_{\rm traj} = 2.0 \times 10^{5}$ realizations. 
Each gray curve corresponds to one sample path of the stochastic realization, while the analytical variance, $\expval*{ [ \delta (\tau) ]^{2} }$, depicted in the wine-red thick curve, and shown on the right axis, characterizes the typical amplitude of the fluctuations as a function of $\tau$. 
The top two panels are for the Fourier-space top-hat (left) and the Gaussian (right) window function, whereas the real-space top-hat window function with the transfer function accommodated is used in the bottom left and right panels, with $\mathcal{P}_{0} = 0.1$ and $\mathcal{P}_{0} = 1.0$ respectively, for later comparison. 

It can be seen in the top-left panel that, though it was already studied in Refs.~\cite{Kushwaha:2025zpz, Saito:2025sny}, the driving noises are the uncorrelated ones in the regime $\widetilde{\tau} \lesssim \widetilde{k}_{\star}$, while after passing the fiducial scale $\widetilde{k}_{\star}$ the trajectories are absorbed towards $\delta (\tau) = 0$ due to the correlation across scales. 
Except for the presence of the uncorrelated nature of the noise, those overall behaviors can also be seen in all the panels, \textit{i.e.}~regardless of the window function used. 
Contrary to the top-left panel, the stochastic trajectories are more organized when either a smooth window function is implemented. 
This behavior renders the degeneracy of the probabilities defined in Section~\ref{subsec:fprob} vanish.
In addition, the oscillatory feature passing down from $\widetilde{W}_{\rm eff} (z)$ can be observed in the bottom panels. 
While the corresponding scale is totally fixed by $\widetilde{k}_{\star}$, the maximum value of $\expval*{ [ \delta (\tau) ]^{2} }$ differs according to the choice of 
the amplitude of the power spectrum $\mathcal{P}_{0}$.  
It should also be noted that the scale at which the variance $\expval*{ [ \delta (\tau) ]^{2} }$ becomes maximum 
depends on the choice of the window function.
For $\widetilde W_{\Theta}$, the variance takes the maximum at the peak position of the power spectrum $\widetilde{\tau} \simeq \widetilde{k}_{\star} = 8$, while the peak positions are shifted to the larger scale (corresponding to smaller $\widetilde{\tau}$) for the smooth window functions. 

\subsection{Formation probabilities} 
\label{subsec:fprob}

At a fixed coarse-graining scale $\tau$, we define the cumulative probability as the probability that a trajectory has crossed the collapse threshold at least once before the coarse-graining scale reaches $\tau$. 
We decompose this event into two mutually exclusive cases (see, \textit{e.g.}, Ref.~\cite{Kushwaha:2025zpz}): 
\begin{align}
    \beta (\tau)
    &\equiv P_1(\tau)+P_2(\tau)
    \notag\\
    &\equiv
    \text{Prob} 
    [ 
        \delta (\tau) > \delta_{\rm c} 
    ] 
    +
    \text{Prob} 
    [ 
        \delta (\tau) \le \delta_{\rm c}
        \ \cap \
        \exists \, \tau' < \tau 
        \ \text{such that} \
        \delta (\tau') > \delta_{\rm c} 
    ] 
    \,\, . 
    \label{eq:beta_P1P2}
\end{align}
The probability $P_1  (\tau)$ counts trajectories which exceed the threshold at the scale $\tau$, whereas $P_2  (\tau)$ counts trajectories that are below the threshold at the scale $\tau$ but have already crossed it at a larger scale (\textit{i.e.}, earlier time). 
Therefore, adding $P_1  (\tau)$ and $P_2  (\tau)$ does not double-count trajectories. 
We note that $P_1  (\tau)$ is the probability that $\delta (\tau)$ exceeds the threshold at a fixed scale. 
Indeed, for a zero-mean Gaussian density contrast with its variance, we have 
\begin{align}
    P_{1} (\tau)
    =
    P [ \delta (\tau) > \delta_{\rm c} ] 
    &=
    \int_{\delta_{\rm c}}^{\infty}
    \frac{\dd\delta}{ \sqrt{2 \pi \langle[ \delta (\tau) ]^{2} \rangle}}
    \exp \bigg( 
        - \bigg\{ 
            \frac{\delta_c}{\sqrt{2\,\left\langle[\delta(\tau)]^2\right\rangle}}
        \bigg\}^{2} 
    \bigg) \notag 
    \\
    &=
    \frac{1}{2}
    \operatorname{erfc}
    \bigg[
              \frac{\delta_{\rm c}}{\sqrt{2 \langle[ \delta (\tau) ]^{2} \rangle}} 
    \bigg]     
    \equiv 
    \beta_{\rm Carr} (\tau) 
    \,\, .
    \label{eq:P1Carr}
\end{align}
Thus, Carr's analytic expression should be identified with $P_1(\tau)$, not with the total excursion-set probability $\beta (\tau) = P_{1} (\tau) + P_{2} (\tau)$. 
It should also be noted that, in the standard excursion-set context, the relation $P_{1}  (\tau) = P_{2}  (\tau)$ \textit{exactly} holds for $ \widetilde W_{\Theta}$, thereby giving the ``factor two'' that resolves the cloud-in-cloud issue. 
This, as we will see soon, ceases to be the case when a smooth window function is used, or when the excursion-set method is applied in the context of PBHs, even for the discontinuous window function. 
The absence of the relation $P_{1} (\tau) = P_{2}  (\tau)$ stems from the non-Markovianity of the stochastic process, and to see how those probabilities contribute to the mass distribution, one is led to conduct numerical simulations, as this article does. 

\begin{figure}
  \centering
  \begin{minipage}{0.46\linewidth}
    \centering
    \includegraphics[width=\linewidth]{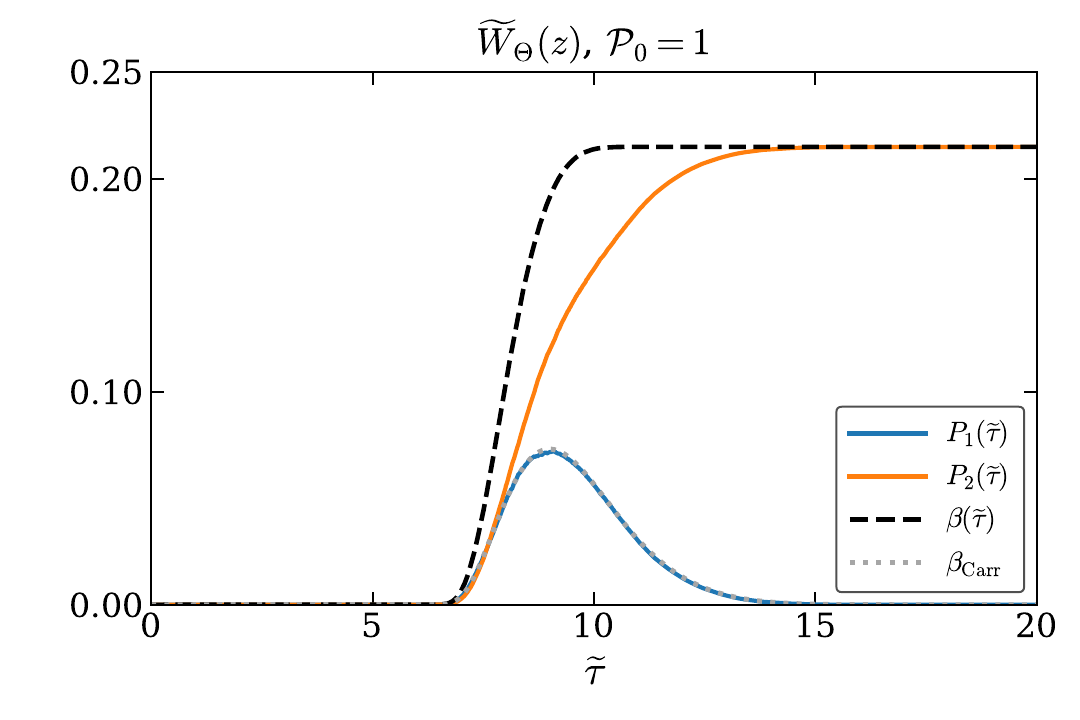}
    \label{fig:p1}
  \end{minipage}
  \hspace{5mm}
  \begin{minipage}{0.46\linewidth}
    \centering
    \includegraphics[width=\linewidth]{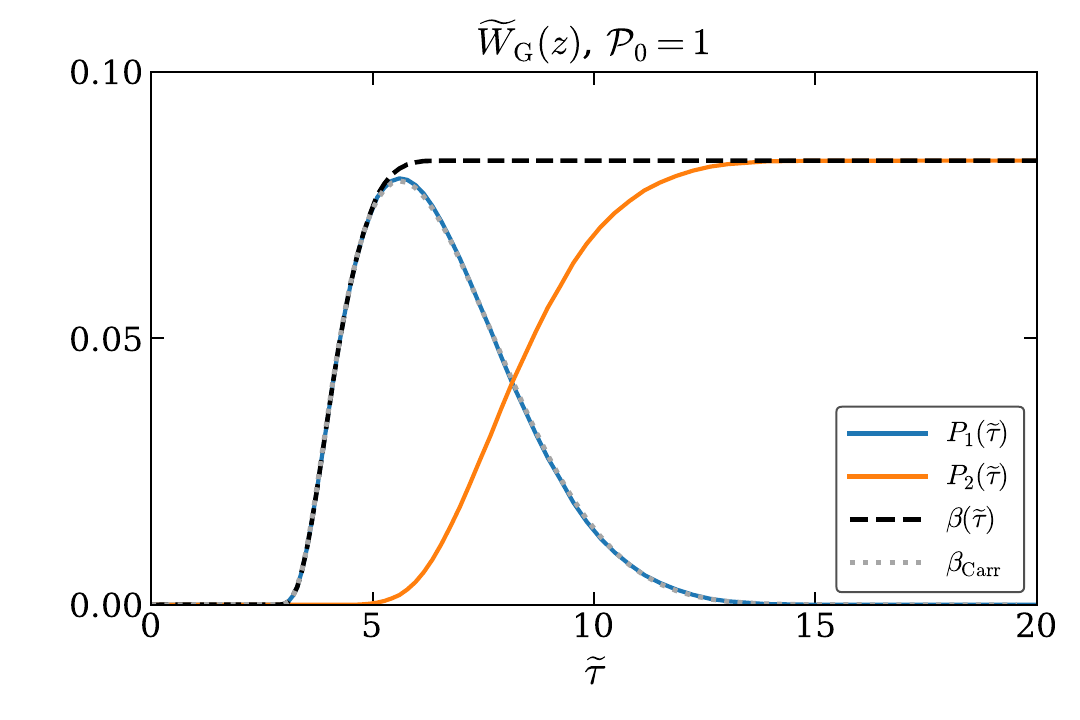}
    \label{fig:p2}
  \end{minipage}
  \vspace{2.5mm}
  \begin{minipage}{0.46\linewidth}
    \centering
    \includegraphics[width=\linewidth]{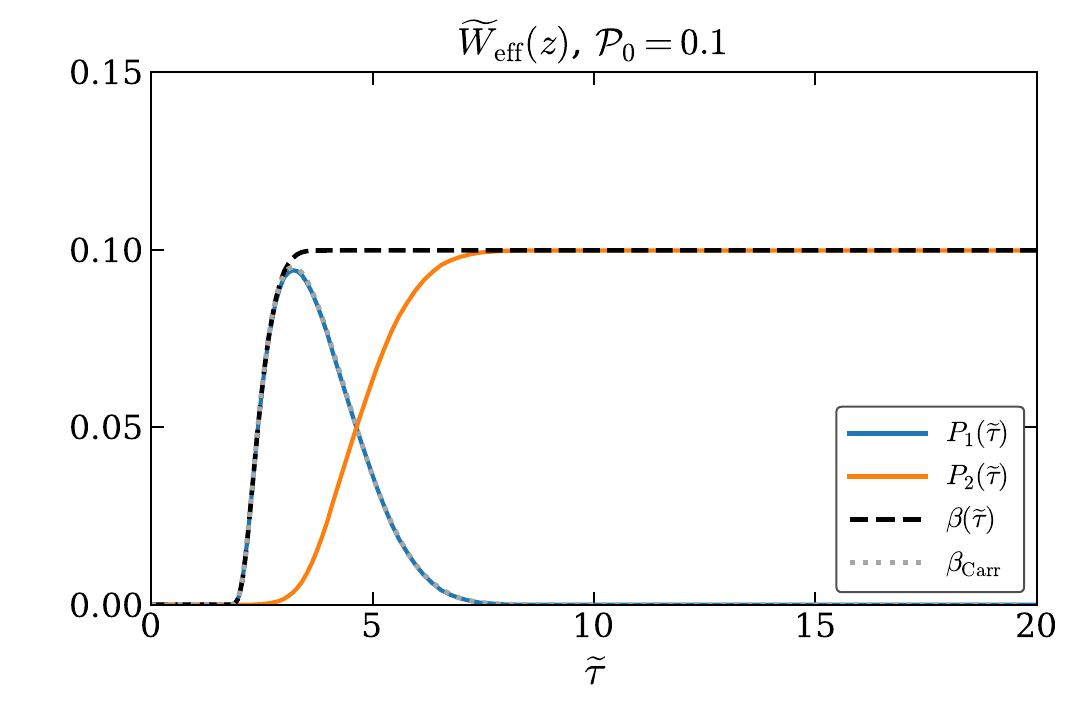}
    \label{fig:p3-1eM1}
  \end{minipage}
  \hspace{5mm}
  \begin{minipage}{0.46\linewidth}
    \centering
    \includegraphics[width=\linewidth]{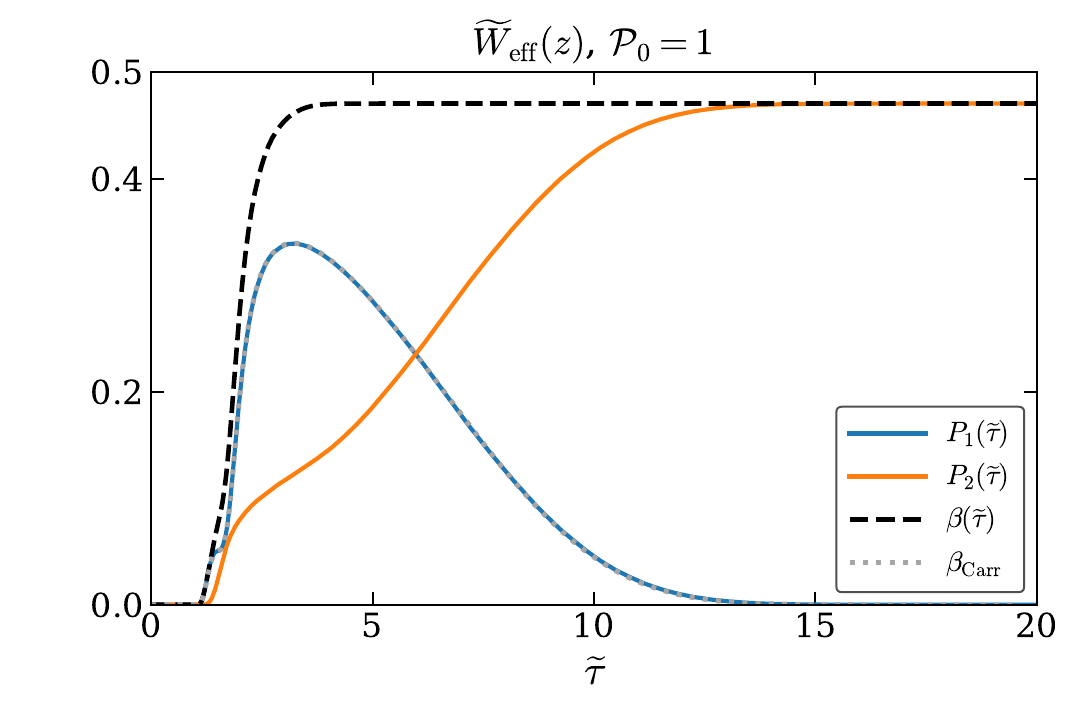}
    \label{fig:p3-1e0}
  \end{minipage}
  \caption{
    Excursion-set probabilities for the three kinds of window functions, Fourier-space top-hat (\textit{top left}), Gaussian (\textit{top right}), real-space top-hat (\textit{bottom}). 
    The amplitudes of the power spectrum, $\mathcal{P}_{0}$, are different between the bottom two panels. 
    The probabilities based on Carr's formula~(\ref{eq:P1Carr}) are shown by the gray dotted curve. 
  }
  \label{fig:prob}
\end{figure}

For the numerically generated stochastic trajectories in Section~\ref{sec:noise}, Figure~\ref{fig:prob} shows the resulting probabilities, in which Carr's formula (\ref{eq:P1Carr}) is also shown for comparison. 
As expected, the numerical blue curve that counts the threshold-piercing events, $P_1(\tau)$, is identical to Carr's analytic probability $\beta_{\rm Carr}(\tau)$, which is expressed as gray dotted curves. 
On the other hand, in all cases, $P_{2} (\tau)$ becomes relevant toward larger $\widetilde{\tau}$, or equivalently toward smaller scale. 
On the degeneracy of the two probabilities, it can only be observed in the case with $ \widetilde W_{\Theta} (z)$ as long as $\widetilde{\tau} \lesssim k_{\star}$, which is consistent with the top-left panel in Figure~\ref{fig:straj}. 
This degeneracy vanishes once a smooth window function is used, as can be seen in the top-right and bottom panels in Figure~\ref{fig:prob}, as was pointed out in Ref.~\cite{Saito:2025sny} for $ \widetilde W_{\rm G} (z)$. 

\begin{figure}
  \centering
  \begin{minipage}{0.46\linewidth}
    \centering
    \includegraphics[width=\linewidth]{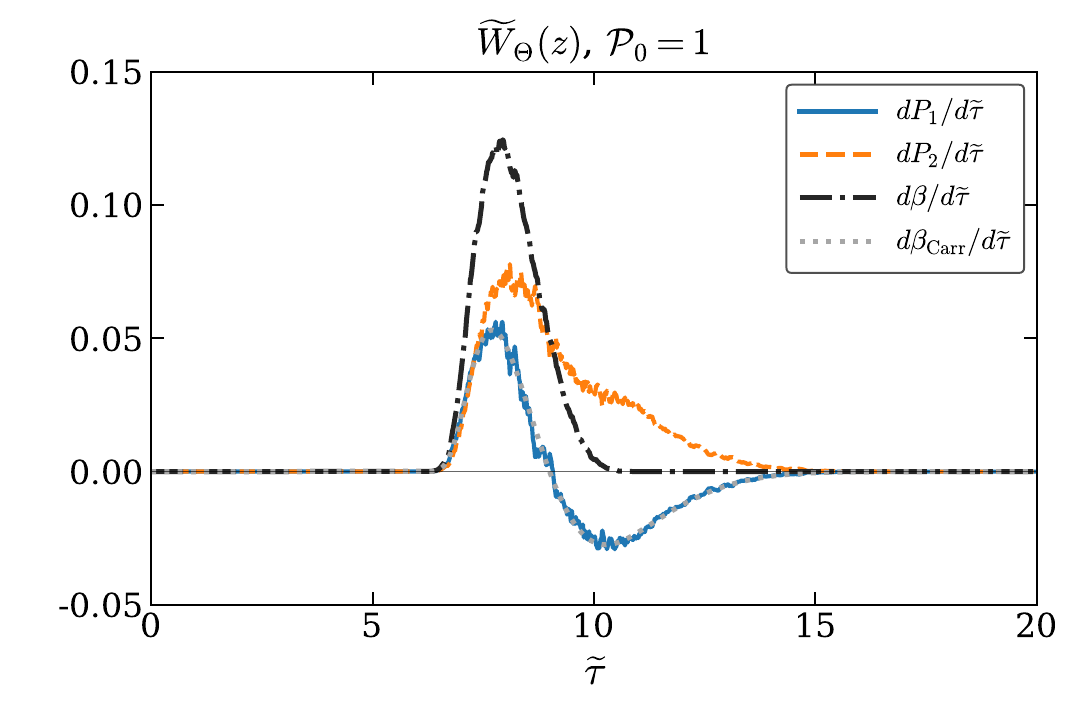}
    \label{fig:dp1}
  \end{minipage}
  \hspace{5mm}
  \begin{minipage}{0.46\linewidth}
    \centering
    \includegraphics[width=\linewidth]{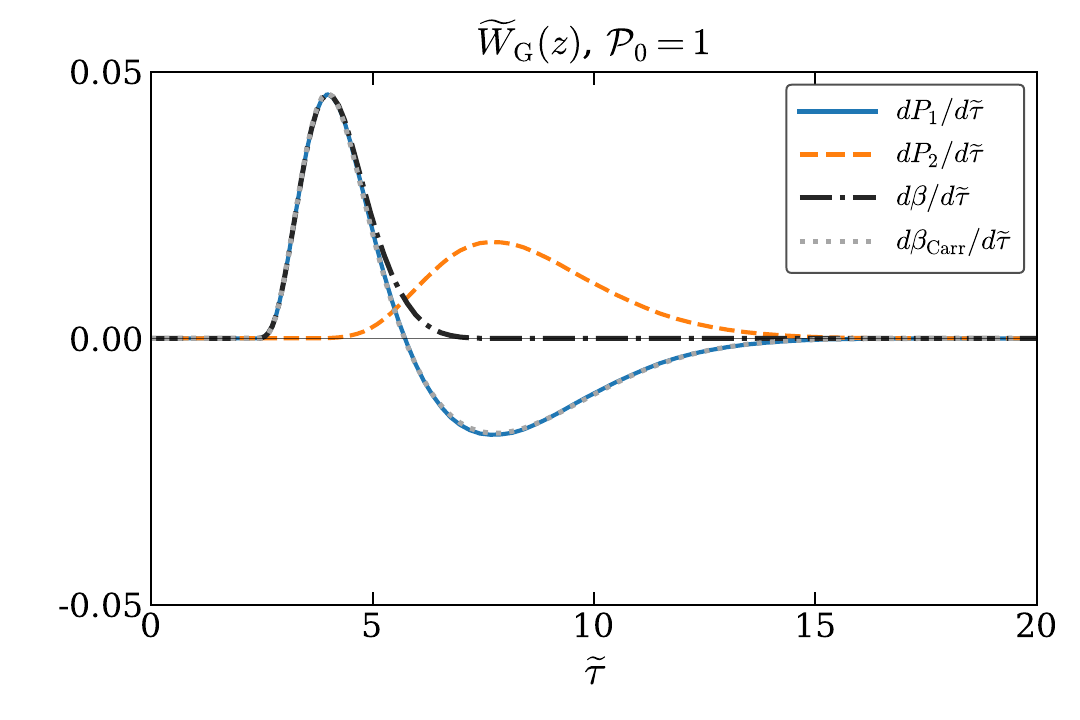}
    \label{fig:dp2}
  \end{minipage}
  \vspace{2.5mm}
  \begin{minipage}{0.46\linewidth}
    \centering
    \includegraphics[width=\linewidth]{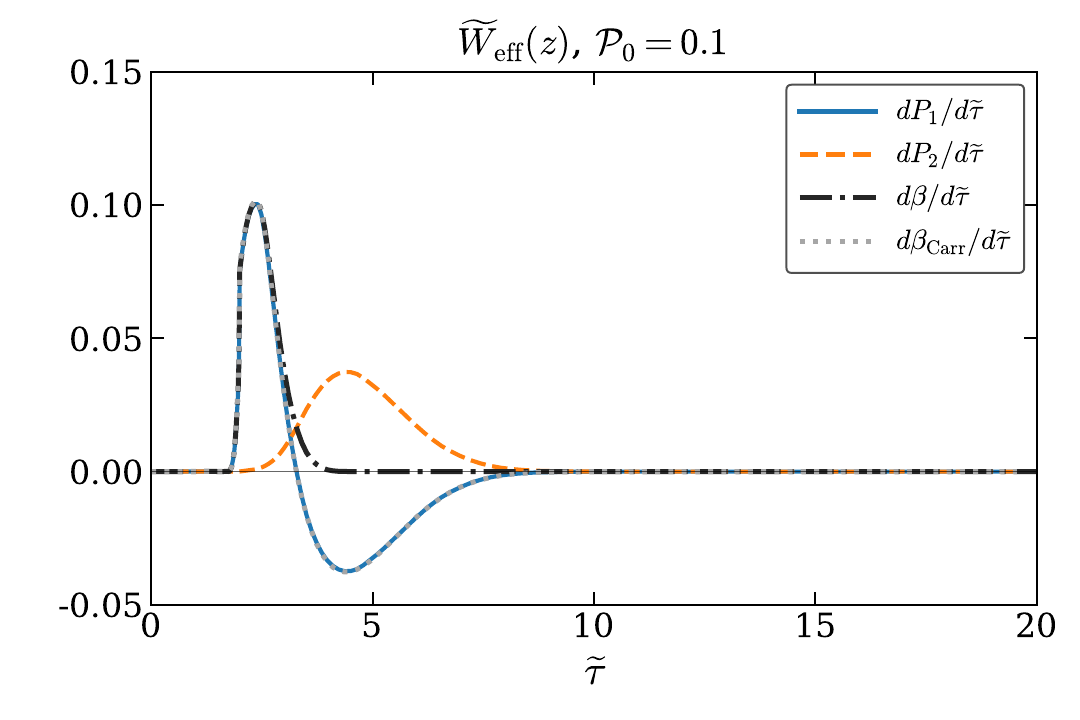}
    \label{fig:dp3-1eM1}
  \end{minipage}
  \hspace{5mm}
  \begin{minipage}{0.46\linewidth}
    \centering
    \includegraphics[width=\linewidth]{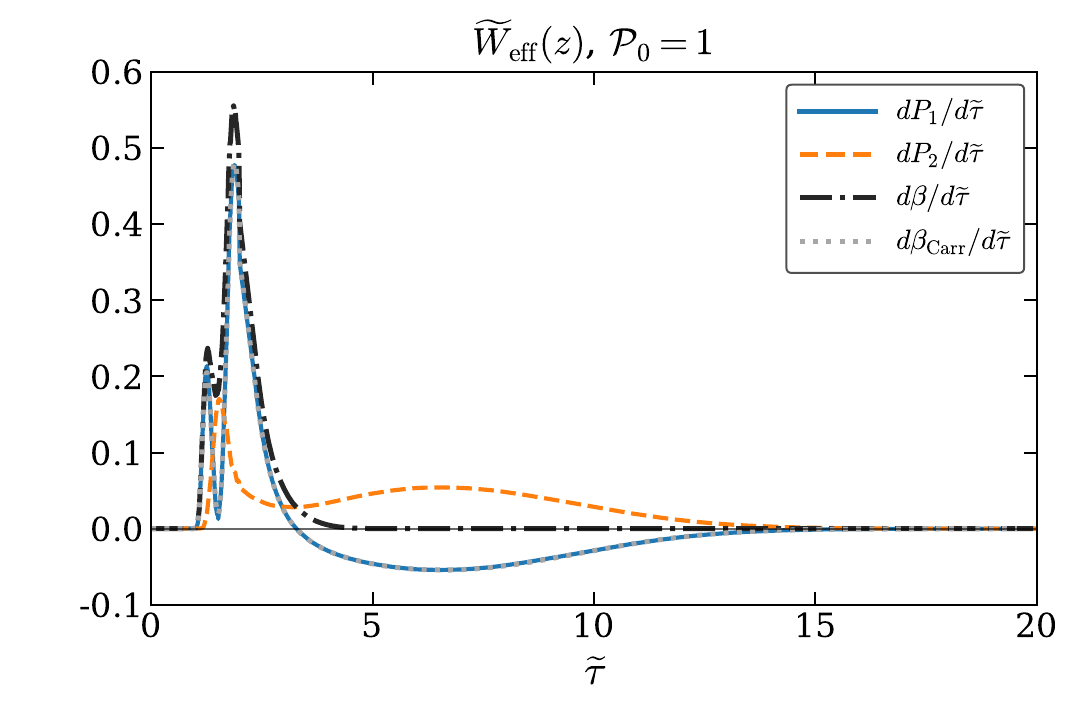}
    \label{fig:dp3-1e0}
  \end{minipage}
  \caption{
    Derivatives of the excursion-set probabilities for the three kinds of window functions, Fourier-space top-hat (\textit{top left}), Gaussian (\textit{top right}), and real-space top-hat (\textit{bottom}). 
    The amplitudes of the power spectrum, $\mathcal{P}_{0}$, are different between the bottom two panels. 
    The derivative of the probabilities based on Carr's formula~(\ref{eq:P1Carr}) is shown by the gray dotted curve.
  }
  \label{fig:dprob}
\end{figure}

The derivatives of the probabilities with respect to $\tau$ are shown in Figure~\ref{fig:dprob}. 
Since $P_{1} (\tau)$ and $P_{2} (\tau)$ have different meanings, their derivatives also inherit the distinct structures, and so does the derivative of the total excursion-set probability,
\begin{equation}
    \frac{\dd \beta (\tau)}{\dd \tau} 
    = 
    \frac{\dd P_{1} (\tau)}{\dd \tau} + \frac{\dd P_{2} (\tau)}{\dd \tau} 
    \,\, . 
\end{equation}
The analytic derivative $\dd \beta_{\rm Carr} (\tau) / \dd \tau$ is, by definition, along with $\dd P_{1} (\tau) / \dd \tau$, which is also seen in Figure~\ref{fig:prob}. 
At the same time, the non-vanishing contribution of $P_2 (\tau)$ shows that the whole probability, $\beta (\tau)$, cannot be constructed by $\beta_{\rm Carr} (\tau)$ alone. 
In particular, for the large-$\tau$ region, the negative value of $\dd P_{1} / \dd \tau$, or $\dd \beta_{\rm Carr} / \dd \tau$, is compensated by $\dd P_2/\dd\tau$, so that the derivative of the whole probability within the excursion-set method remains non-negative, as can be seen in the figure. 
This compensation can be found regardless of the choice of the window function, and highlights the role of $P_{2} (\tau)$ in addition to $P_{1} (\tau) = \beta_{\rm Carr} (\tau)$. 
It was already found for $\widetilde{W}_{\Theta} (z)$ in Ref.~\cite{Kushwaha:2025zpz}, for $\widetilde{W}_{\rm G} (z)$ in Ref.~\cite{Saito:2025sny}, and it is now found that the same holds also when the real-space top-hat window function is used. 
Therefore, in summary, the contribution of $P_{2} (\tau)$ is not negligible in terms of the non-negativity of $\dd \beta (\tau) / \dd \tau$, for all the window functions typically used in the literature. 

\section{Mass function of PBHs}
\label{massfunction}

The formation probabilities as a function of $\widetilde{\tau}$ have been extracted from the stochastic trajectories in Section~\ref{stochastic}. 
Based on those probabilities, this section derives the mass function of PBHs.
We shall begin with giving the mass-scale relation in Section~\ref{subsec:relm}. 
By contrasting it with the prediction from Carr's formula in Section~\ref{sec:4.2.1}, we clarify the role of $P_{2}(\tau)$ from the viewpoint of the mass function. 
In Section~\ref{sec:mfex}, we continue to discuss the mass function, while turning our focus to the window-function dependence. 

\subsection{Relation between mass and coarse-graining scale}
\label{subsec:relm}

To convert the coarse-graining scale into a mass scale, we use the standard horizon-mass relation during the radiation-dominated era. 
The scaling follows by chasing the thermal history of the universe, and is given by~\cite{Young:2014ana}, 
\begin{equation}
    M_{\rm PBH} (\tau) 
    \simeq 
    \frac{3}{2}M_{\rm H, \, eq}
    \left( 
        \frac{k_{\rm eq}}{\tau} \right 
    )^{2}
    \left( 
        \frac{g_{\ast, \, \rm eq}}{g_{\ast}} 
    \right)^{1/3} 
    \,\, .
    \label{eq:msrel}
\end{equation}
Here, we have identified the coarse-graining scale with the horizon-entry scale, $\tau = aH = k$. 
The quantities $M_{\rm H, \, eq}$ and $k_{\rm eq}$ are the horizon mass and wavelength scale at the matter-radiation equality, respectively, and are used as reference quantities. 
$g_{\ast}$ is the number of relativistic degrees of freedom and $g_{\ast, \, \rm eq}$ denotes its value at the matter-radiation equality.

In terms of the dimensionless variables, Eq.~(\ref{eq:msrel}) becomes
\begin{equation}
\frac{M_{\rm PBH}}{M_\odot}
\simeq
\frac{3}{2\widetilde{\tau}^{2}}
\frac{M_{H,\rm eq}}{M_\odot}
\left(\frac{k_{\rm eq}}{\Delta}\right)^2
\left(\frac{g_{\ast,\rm eq}}{g_\ast}\right)^{1/3}.
\end{equation}
We shall use the numerical data, $M_{\rm H, \, eq} / M_{\odot} \approx 3.5 \times 10^{17}$ and $( g_{\star, \, {\rm eq}} / g_{\star} )^{1/3} \approx 0.3$, following Refs.~\cite{Sasaki:2018dmp,Carr:2020gox,Carr:2020xqk}, to get 
\begin{equation}
    \qty( \frac{ \Delta }{ 10^{8} k_{\rm eq} } )^{2} 
    \frac{M_{\rm PBH}}{ M_{\odot} } 
    \approx \frac{16}{\widetilde{\tau}^{2}} 
    \,\, . 
    \label{eq:msrel_non_dim}
\end{equation}
For our convenience, we neither specify a concrete value of $\Delta$ nor restore the dimension of $\tau$. 

The probability $\beta (\tau)$ as a function of $\tau$ can be converted to the probability as a function of $\ln M_{\rm PBH}$, by virtue of the relation (\ref{eq:msrel_non_dim}), through 
\begin{equation}
    \ln M_{\rm PBH} 
    = 
    - 2 \ln \tau + \mathrm{const} 
    \,\, ,
    \qquad
    \left| 
        \frac{\dd \tau}{\dd \ln M_{\rm PBH}} 
    \right| 
    = 
    \frac{\tau}{2} 
    \,\, .
\label{eq:jacobian_tau_lnM}
\end{equation}
Accordingly, the derivatives of the probabilities are related by 
\begin{equation}
    \frac{\dd \beta (M_{\rm PBH})}{\dd \ln M_{\rm {PBH}}}
    = 
    \left| 
        \frac{ \dd \tau }{ \dd \ln M_{\rm PBH} } 
    \right| 
    \frac{ \dd \beta (\tau) }{ \dd \tau } 
    = 
    \frac{\widetilde{\tau}}{2} \dv{ \beta (\tau) }{ \widetilde{\tau} } 
    \,\, .
    \label{eq:jacobian_general_abs}
\end{equation}
In Eq.~(\ref{eq:jacobian_general_abs}), the left-hand side should be understood as a function of $M_{\rm PBH}$, while the right-hand side is a function with respect to $\widetilde{\tau}$, which is defined in Eq.~(\ref{eq:beta_P1P2}). 
Re-expressing the right-hand side of Eq.~(\ref{eq:jacobian_general_abs}) as a function of $M_{\rm PBH}$, using the relation (\ref{eq:msrel}), the mass distribution can be obtained for each curve displayed in Figure~\ref{fig:dprob}. 

\subsection{Mass function in excursion-set method}
\label{subsec:esc}

We shall use the relation between the coarse-graining scale $\widetilde{\tau}$ and the horizon mass to express the mass function in logarithmic-mass units, as a function of $M_{\rm PBH}$ itself instead of $\widetilde{\tau}$, see Eq.~(\ref{eq:jacobian_general_abs}). 

\subsubsection{Comparison with Carr's formula}
\label{sec:4.2.1}
\begin{figure}
  \centering
  \begin{minipage}{0.46\linewidth}
    \centering
    \includegraphics[width=\linewidth]{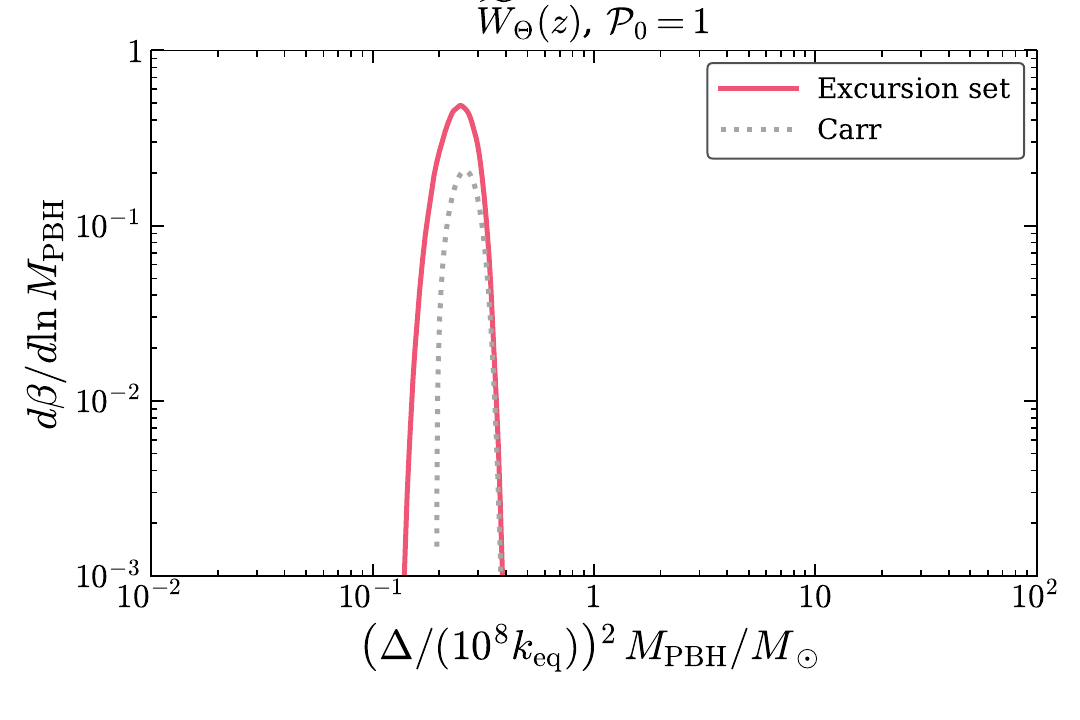}
    \label{fig:mf1}
  \end{minipage}
  \hspace{5mm}
  \begin{minipage}{0.46\linewidth}
    \centering
    \includegraphics[width=\linewidth]{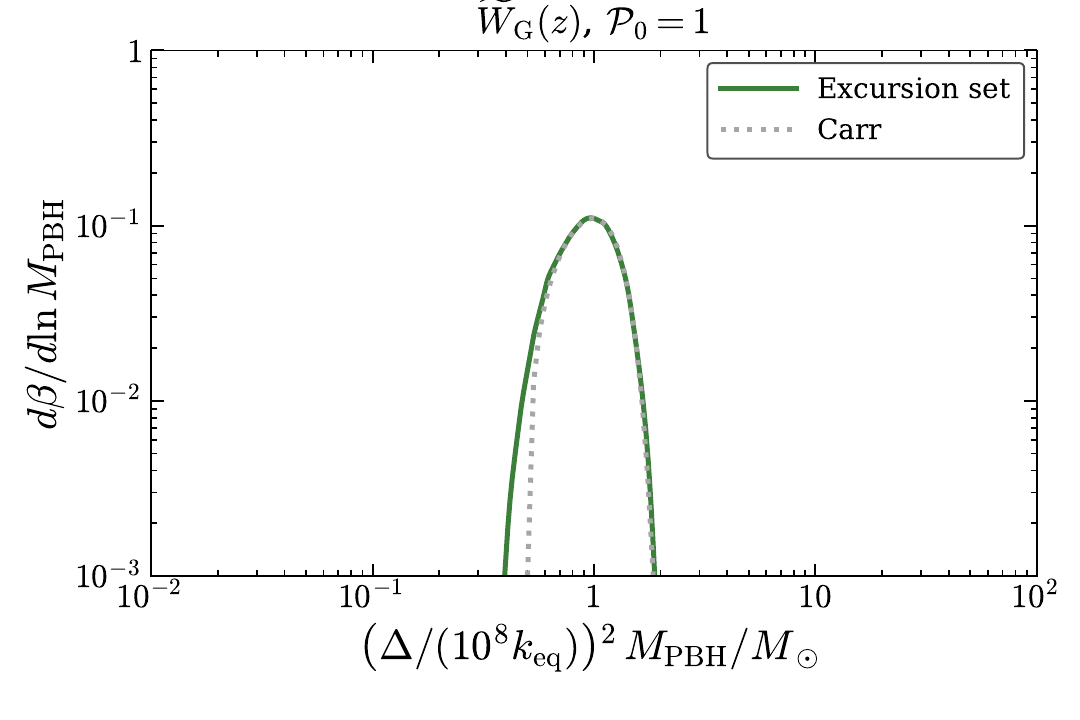}
    \label{fig:mf2}
  \end{minipage}
  \vspace{2.5mm}
  \begin{minipage}{0.46\linewidth}
    \centering
    \includegraphics[width=\linewidth]{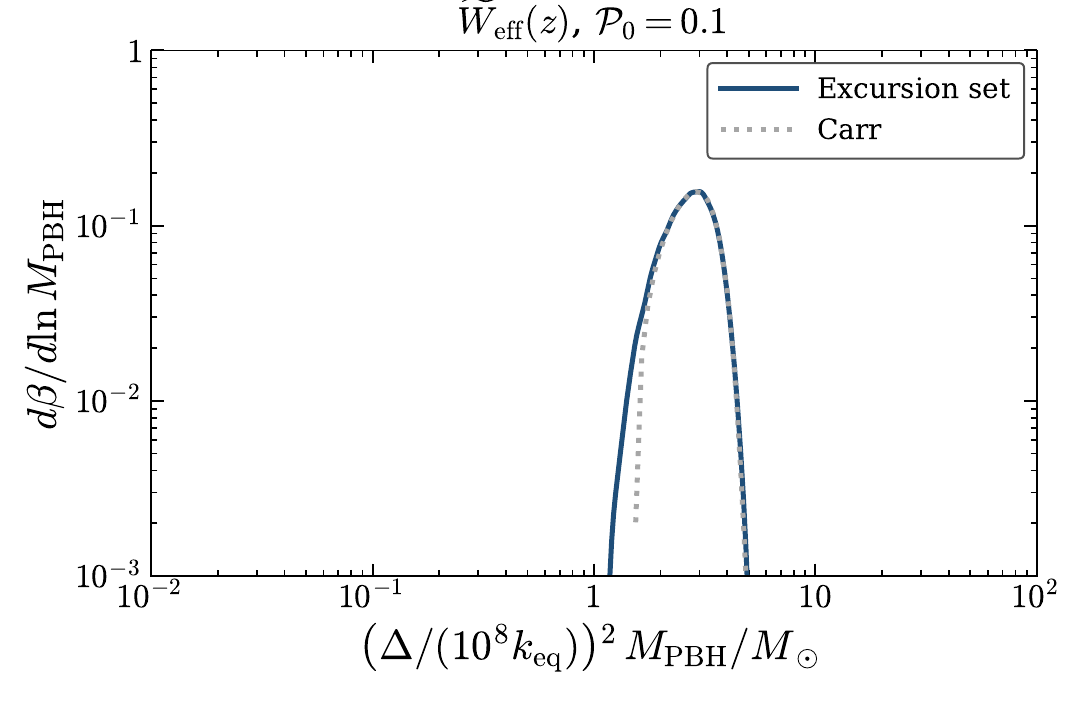}
    \label{fig:mf3-1eM1}
  \end{minipage}
  \hspace{5mm}
  \begin{minipage}{0.46\linewidth}
    \centering
    \includegraphics[width=\linewidth]{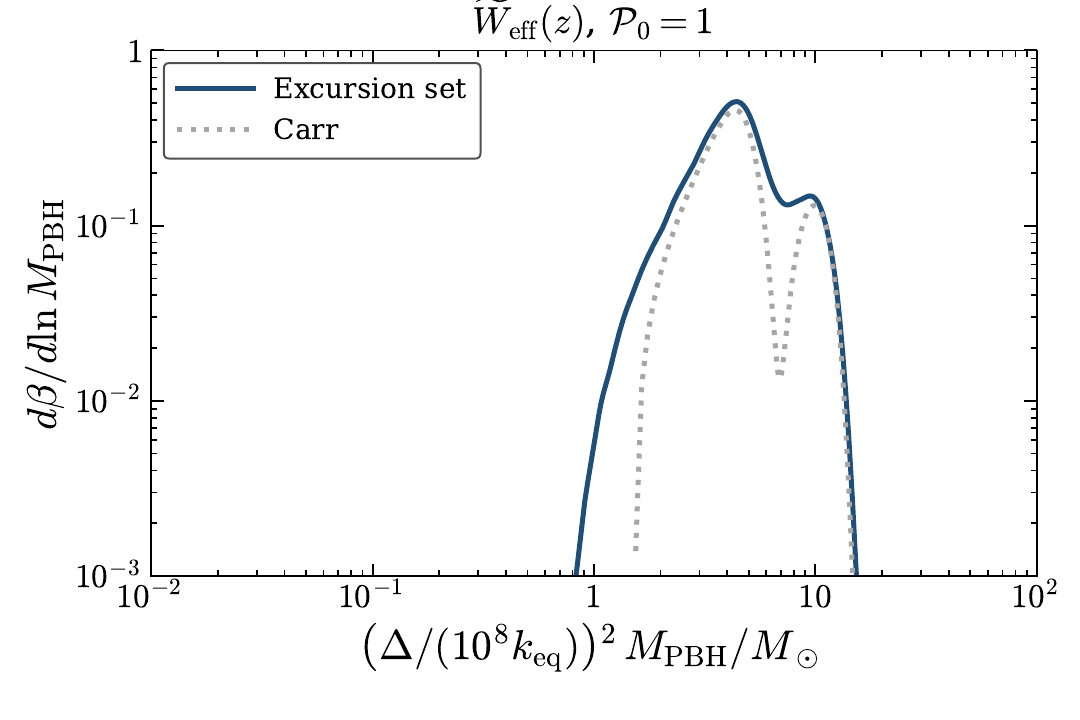}
    \label{fig:mf3-1e0}
  \end{minipage}
  \caption{
    The derivative mass fractions of PBHs formed from the density contrast, coarse-grained by the three kinds of window functions, Fourier-space top-hat (\textit{top left}), Gaussian (\textit{top right}), and real-space top-hat (\textit{bottom}). 
    The amplitudes of the power spectrum, $\mathcal{P}_{0}$, are different between the bottom two panels. 
    The dotted curves denote the mass function computed from Carr's formula~(\ref{eq:carrmassfuncvariance}).}
  \label{fig:fbeta}
\end{figure}

Figure~\ref{fig:fbeta} 
contrasts the mass function predicted by the excursion-set method, taking both probabilities $P_{1} (\tau)$ and $P_{2} (\tau)$ into account, with the estimation based on Carr's formula. 
We numerically reconstructed the mass function, $\dd \beta (M_{\rm PBH}) / \dd \ln M_{\rm PBH}$, from the ensemble of first threshold crossings and normalized so that its integral reproduces the total formation fraction, $\beta (M_{\rm PBH})$. 
Here, $\beta_{\rm Carr} (\tau)$ is defined in Eq.~(\ref{eq:P1Carr}) as a function of $\tau$, and we convert its $\tau$-derivative to a derivative with respect to $\ln M_{\rm PBH}$ by means of Eq.~\eqref{eq:jacobian_general_abs}, which reads 
\begin{equation}
    \frac{\dd \beta_{\rm Carr} (M_{\rm PBH})}{\dd \ln M_{\rm PBH}}
    = 
    \frac{\tau}{2 \sqrt{\pi}} 
    \frac{\dd}{\dd \tau} 
    \bigg\{ 
        \frac{\delta_{\rm c}}{\sqrt{2 \langle[ \delta (\tau) ]^{2} \rangle}} 
    \bigg\} 
    \exp \bigg( 
        - \bigg\{
            \frac{\delta_{\rm c}}{\sqrt{2 \, \langle[ \delta (\tau)]^{2} \rangle}} 
        \bigg\}^2 
    \bigg) 
    \,\, . 
    \label{eq:carrmassfuncvariance}
\end{equation}
It should be noted that, as is observed in Figure~\ref{fig:dprob}, the left-hand side of Eq.~(\ref{eq:carrmassfuncvariance}) is not positive for all $\tau$, contrary to the total probability $\beta ( \tau ) = P_{1} (\tau) + P_{2} (\tau)$. 
For this reason, only the positive region of the quantity $\dd \beta_{\rm Carr} (M_{\rm PBH}) /\dd \ln M_{\rm PBH}$ is used for our purposes here. 

From the top-left, top-right, and bottom-left panels in Figure.~\ref{fig:fbeta}, it can be seen that both $\beta_{\rm Carr} (M_{\rm PBH})$ and $\beta (M_{\rm PBH})$ are in agreement especially \textit{around the maximum} for all the window functions but $\widetilde{W}_{\Theta} (z)$, \textit{i.e.}~for the smooth window functions, whereas a difference can be seen in the low-mass tails. 
The reason why the total mass function consists only of $P_{1}(\tau)$ for the two window functions, as long as one is interested around the maximum, is that the two formation probabilities are not in degeneracy. 
Therefore, one of the statements made in Ref.~\cite{Saito:2025sny} that the degeneracy renders $P_{2}(\tau)$ irrelevant \textit{except the low-mass tail} has been confirmed, including the case with the real-space top-hat window function. 
In the bottom-right panel, although the presence of the secondary maximum comes from the oscillatory feature of $\widetilde{W}_{\rm eff} (z)$, the same statement remains valid that the abundance estimated within the excursion-set method and $\beta_{\rm Carr} (M_{\rm PBH})$ are get along with each other in the vicinity of the characteristic scale, at which $\dd \beta (M_{\rm PBH}) / \dd \ln M_{\rm PBH}$ becomes maximum. 
The role of $P_{2} (\tau)$ is also reflected in the mass dependence of the deviation from Carr's estimate. 
Since smaller PBH masses correspond to larger $\widetilde{\tau}$, the small-mass side is reached only after the trajectories have already sampled large-mass scales, by definition of $P_{2} (\tau)$. 
This formation channel cannot be captured by $P_{1} (\tau) = \beta_{\rm Carr} (\tau)$. 
For the Fourier-space top-hat window function, on the other hand, the maximum value of the mass function cannot be estimated correctly solely by $\beta_{\rm Carr}(\tau)$, since $P_{1}(\tau) \approx P_{2}(\tau)$ within the range of $\tau$ where the derivatives of those probabilities grow, see the top-left panel in Figure~\ref{fig:prob}. 

In summary, both $P_{1} (\tau)$ and $P_{2} (\tau)$ need to be accommodated to ensure the non-negativity of the mass function, as was discussed in Ref.~\cite{Kushwaha:2025zpz}. 
The estimation based on Carr's formula can nevertheless be used as long as one is interested in the scale at which the mass function is maximized, only when the density contrast is coarse-grained by a smooth window function. 
In other words, the formation probabilities based on Carr's estimate and the excursion-set method coincide with each other near the characteristic mass scale, while the excursion-set method determines the whole mass function including its low-mass tail. 
All these features come from the continuity of the window functions in Fourier-space, by which the stochastic noises are completely correlated across scales and the two probabilities, $P_{1} (\tau)$ and $P_{2} (\tau)$, behave differently. 
Contrary to this, neglect of $P_{2} (\tau)$ cannot be justified even around the maximum of the mass function when the discontinuous Fourier-space window function is implemented. 

\subsubsection{Comparison among window functions}
\label{sec:mfex}

\begin{figure}
  \centering
  \includegraphics[width=0.82\linewidth]{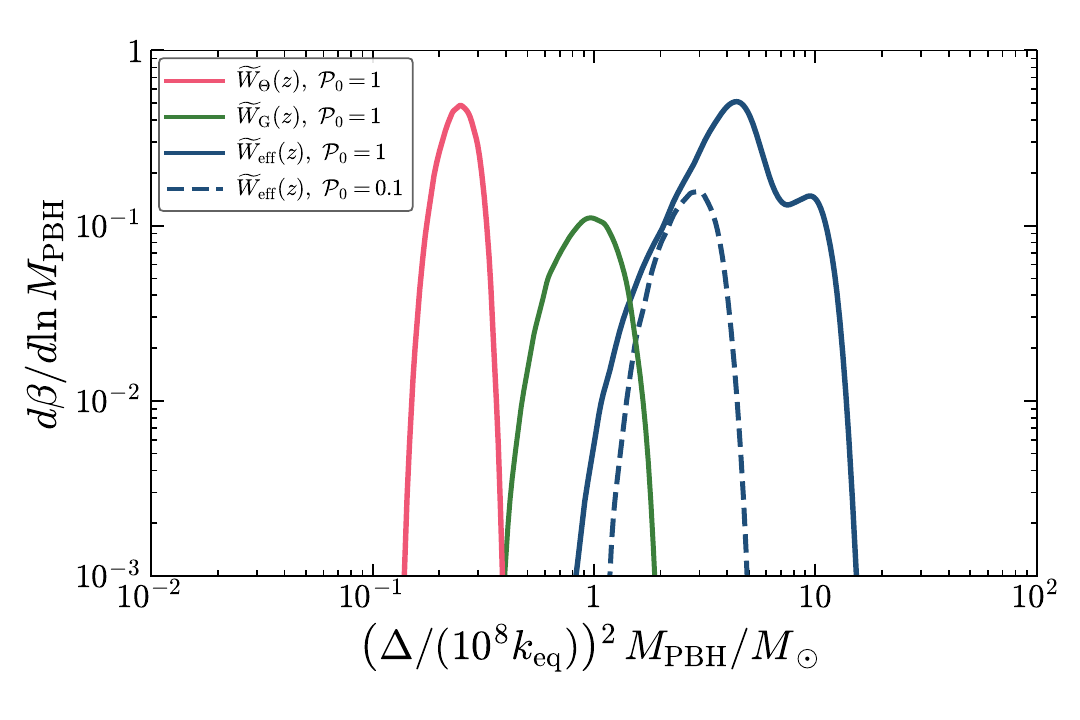}
  \caption{Comparison of numerical mass functions obtained with different coarse-graining windows. The parameters are the same as Eq.~(\ref{eq:params}).}
  \label{fig:massfunc_carr_overlay_added} 
\end{figure}

Figure~\ref{fig:massfunc_carr_overlay_added} shows the mass functions for the three window functions, \textit{i.e.}~the colored curves shown in Figure~\ref{fig:fbeta} in a single plane, to contrast the effect of the choice of the window function. 
The colors correspond to those in Figure~\ref{fig:fbeta}. 
As was done there, the amplitude $\mathcal{P}_{0}$ of the power spectrum is fixed to be $\mathcal{P}_{0} = 1.0$ for the Fourier-space top-hat and Gaussian window functions, whereas for the real-space top-hat window function with the transfer function $\mathcal{P}_{0} = 0.1$ and $1.0$ are contrasted. 

One immediately notices that all the characteristic quantities in the mass function, such as the location of the maximum, are sensitive to the window function. 
The scale at which $\dd \beta (M_{\rm PBH}) / \dd \ln M_{\rm PBH}$ is maximized for $\widetilde{W}_{\Theta} (z)$ corresponds to the fiducial scale, $\widetilde{k}_{\star}$, at which $\mathcal{P}_{\zeta} (k)$ is also maximized, and which is almost directly passed down to the mass function. 
On the other hand, since a smooth window function effectively shifts the scale, at which the variance of $\delta (\tau)$ is maximized, see also the thick curves in Figure~\ref{fig:straj}, the corresponding scale of the mass function is also shifted, towards a heavier mass than that corresponds to $\widetilde{k}_{\star}$. 
The overall shape of $\dd \beta (M_{\rm PBH}) / \dd \ln M_{\rm PBH}$ is also sensitive to the choice of the window function. 
As was mentioned in Section~\ref{sec:4.2.1}, the double-maximum structure of the curve for $\widetilde{W}_{\rm eff} (z)$ with $\mathcal{P}_{0}=1$ in Figure~\ref{fig:massfunc_carr_overlay_added} comes from the oscillatory structure of the window function. 
Focusing on the other three curves, each of them consists of a single maximum at the scale discussed above. 
However, the way $P_{1} (\tau)$ and $P_{2} (\tau)$ contribute to the total mass function is different between $\widetilde{W}_{\Theta} (z)$ and the other two window functions. 
This is because, as can be seen in Figure~\ref{fig:dprob}, the derivatives of those probabilities behave differently once the smooth window function is used. 
For the case with $\widetilde{W}_{\Theta} (z)$, the degeneracy of $P_{1} (\tau)$ and $P_{2} (\tau)$ makes the contributions of them comparably, while the total mass function is determined almost solely by $P_{1}$ for $\widetilde{W}_{\rm G} (z)$, except its smaller-mass tail. 

\section{Conclusion}
\label{conclusion}

In this work, we studied the formation of PBHs based on the excursion-set method, in which colored noises drive the stochastic process that governs the response of the coarse-grained density contrast to the variation of the coarse-graining scale. 
By analytically deriving the covariance matrix of the stochastic noise, $\partial \delta (\tau) / \partial \tau$, and numerically generating the noises with the desired properties, we drew the stochastic realizations from which the threshold-crossing statistics were extracted and the mass function of PBHs was reconstructed. 

The stochastic trajectories generated and observed in Figure~\ref{fig:straj} led to the two kinds of formation probabilities, $P_1(\tau),~P_2(\tau)$, see Figure~\ref{fig:dprob}. 
The Fourier-space top-hat, Gaussian, and real-space top-hat window functions are considered to highlight the common and different features among them. 
On this point, the transfer function is properly taken into account to make the large-$z$ behavior of the window function well-behaved. 
In addition to the degeneracy of those probabilities for $\widetilde{W}_{\Theta} (z)$ in Ref.~\cite{Kushwaha:2025zpz} and non-degeneracy for $\widetilde{W}_{\rm G} (z)$ in Ref.~\cite{Saito:2025sny}, we have confirmed that they are not in degeneracy also for the case with the real-space top-hat window function. 

These trajectories and probabilities then enabled us to reconstruct the mass function of PBHs as a function of their mass instead of the coarse-graining scale. 
We contrasted the mass function within the excursion-set method with the one based on Carr's pioneering formula to clarify their coincidence around the characteristic scale and how their differences arise. 
In other words, in the vicinity of the characteristic scale at which the mass function is maximized, Carr's formula is in agreement with the mass function based on the excursion-set method. 
However, not only Carr's probability but also the probability $P_{2} (\tau)$ must be included in order for $\dd \beta (M_{\rm PBH}) / \dd \ln M_{\rm PBH}$ to be well-behaved, \textit{i.e.}~in order to ensure its non-negativity. 
Also, we investigated the dependence of the mass function on the choice of the window function. 
When one uses the Fourier-space top-hat window function, or when one is interested in the low-mass tail of the mass function with a smooth window function used, the contribution from 
$P_2(\tau)$ plays a non-negligible role. 
In contrast, though the full probability is given by $\beta (\tau) = P_{1} (\tau) + P_{2} (\tau)$ especially in the low-mass (\textit{i.e.}~large $\widetilde{\tau}$) regime, the approximation $\beta (\tau) \approx P_{1} (\tau)$ is of practical use when the estimation is performed around the characteristic scale, as long as a smooth window function is used. 

\section*{Acknowledgments}
We are grateful to Tomohiro Harada and Tsutomu Kobayashi for fruitful discussions. This work was partially supported by Rikkyo University Special Fund for Research (H.I.), the National Research Foundation of Korea Grant funded by the Korean Government RS-2024-00336507 (D.S.), and JSPS Overseas Research Fellowships (K.T.). 

\bibliographystyle{JHEP}
\bibliography{refs}

\end{document}